# Modeling Influence with Semantics in Social Networks: a Survey

*Gerasimos Razis*[1], *Ioannis Anagnostopoulos*[1], *Sherali Zeadally*[2]

[1]Department of Computer Science and Biomedical Informatics, University of Thessaly

{razis, janag}@dib.uth.gr

[2]*College of Communication and Information, University of Kentucky*

*szeadally@uky.edu*

**Abstract**

The discovery of influential entities in all kinds of networks (e.g. social, digital, or computer) has always been an important field of study. In recent years, Online Social Networks (OSNs) have been established as a basic means of communication and often influencers and opinion makers promote politics, events, brands or products through viral content. In this work, we present a systematic review across i) online social influence metrics, properties, and applications and ii) the role of semantic in modeling OSNs information. We found that both areas can jointly provide useful insights towards the qualitative assessment of viral user-generated content, as well as for modeling the dynamic properties of influential content and its flow dynamics.

# 1. Introduction

Nowadays, hundreds of millions of messages are shared on a daily basis among the users of Online Social Networks (OSNs). These users vary from citizens to political persons and from news agencies to large multinational corporations. In this “ocean” of information, a challenging task is the discovery of the important actors who are able to influence others and produce messages of high social quality, importance and recognition. Those influential users are also called *opinion leaders* [15], *domain experts* [16], *influencers* [111], *innovators* [119], *prestigious* [120] or *authoritative actors* [121]. Often, their degree of influence is also complemented or affected by various quality measurements which are based on their social semantics. The latter can either be related to the content of the messages (e.g. keywords, hashtags) or to the metadata of the user (e.g. activity, relationship details).

In this paper, we study two major aspects of OSNs, namely the online social influence (Section 3) and the role of social semantics (Section 4) in OSNs, towards the qualitative assessment of viral user-generated content (Section 5). Specifically, we examine how influence can be measured or predicted and what kinds of methodologies are used to measure influence (e.g. based on topology, diffusion or social authority), and what are the application domains. Regarding the role of semantics in OSNs, we analyze related works based on Semantic Web technologies along with network theory and graph properties for topic identification, detection of similar users and communities, as well as user personalization (e.g. interests, suggestions, and so on).

In order to perform a more detailed analysis and to adequately cover all perspectives of the aforementioned two aspects, we analyzed the reviewed related research works according to the hierarchical classification scheme depicted in Figure 1. In most of the cases, a referred work does not fall with the scope of only one topic, thus demonstrating that related research efforts in these fields are complementary.

More specifically, with respect to the online social influence, we classify the related works according to the following four topics.

- *Topic 1 - Influence Metrics*: This topic includes works proposing methodologies that define online social influence and how to measure it. Thus, this topic is further divided in three subtopics namely a) Direct social information-based metrics, b) Hyperlink-based metrics, and c) Metrics based on machine-learning techniques.

- *Topic 2 - Information Flow and Influence*: This topic examines the impact of influence / influential users with respect to viral properties of information as well as information propagation and information diffusion. Although there is no clear distinction between ‘propagation’ and ‘diffusion’ in the literature covering the OSNs and often these terms are used interchangeably, in this survey we explicitly examine separately the impact of influence in information propagation and information diffusion. Diffusion is about the spread of information from a starting node toward the rest of the network, while propagation takes into consideration the intermediate nodes as well, which receive, process, and further decide whether to re-transmit, re-direct or block the information. Thus, in this work we divide information flow and influence topic into two subtopics namely, propagation-oriented and diffusion-oriented.

- *Topic 3 - Network / Graph Properties*: This category contains works which utilize the topology of a network or its structure in order to measure influence. Usually only a fraction of the whole network is used due to hardware or complexity limitations.
- *Topic 4 - Applications*: This topic presents the usage of the above metrics, mainly in applications that provide solutions for opinion makers, data analysts and information scientists. This topic is further divided into three subtopics namely, a) Ranking, b) Recommendation and c) Other application domains, such as sentiment analysis, event detection, and so on.

As for the role of social semantics in the provision of a qualitative assessment of viral user-generated content, we classify the related works we have reviewed into three topics.

- *Topic 1 - Social Modeling*: This topic contains approaches that adopt semantics for modeling the logical topology and structure of online social networks and media as well as the disseminated information.
- *Topic 2 - Social Matching*: The studies presented on this topic exploit the use of social semantics for identifying similar properties and activities with respect to user-generated content, description of real-life events, as well as revealing user interests and behavioral patterns across different online social media users. Thus, we divide this topic into two subtopics, namely a) User-oriented (e.g. similar user recommendation, user preferences, and so on), and b) Topic and Event-oriented (e.g. topic profiling and user interest, event detection, product marketing).
- *Topic 3 - Community Detection*: This category covers works that use social semantics for the detection of communities in OSNs.

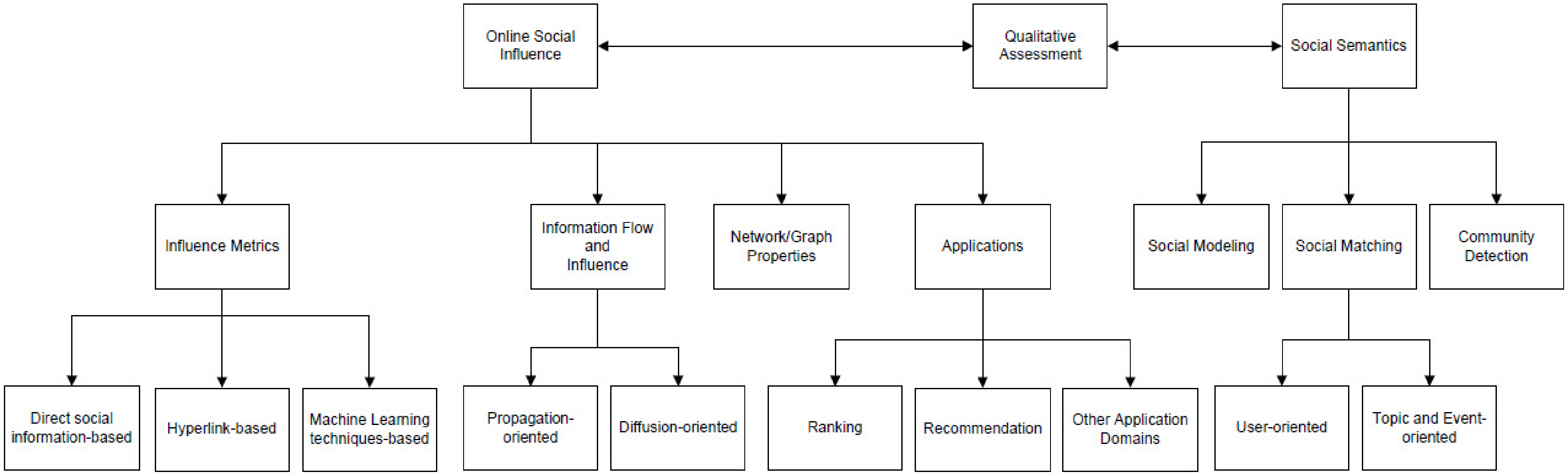

Figure 1: The hierarchical classification scheme followed in this work

This survey aims to help both researchers and data scientists to better understand how viral content is propagated, the role and effect of influential nodes in its diffusion, how we can measure the influence of users in social networks and the reasons why the proper use of semantics for users and their generated contents can provide useful insights and qualitative conclusions for numerous domains such as marketing, information retrieval, recommendation systems, community and/or event detection, query expansion, thematic categorization, homophily tendency and sentiment analysis.

To conduct our literature review, we collected 126 studies strongly related to the aforementioned issues. Initially, we used a specific set of related keywords (some indicative keyword as depicted in Table 1) as input for the discovery of relevant publications by submitting them through the academic digital library and search engine Application Programming Interfaces (APIs). Specifically, we utilized open access repositories (e.g. Google Scholar[1], arXiv[2], SSRN[3]) and digital libraries that request subscription (e.g. ACM Digital Library[4], IEEE Xplore[5], Elsevier[6], others). Next, we performed a review on the selected studies to highlight the most relevant topics and subtopics related to the influence in OSNs and social semantics. In the final step, we have further filtered the selected works based on their date of publications, thus keeping the most recent. However, in order not to exclude the older but significant related works (with high citation counts), we described their impact in the newer works that have cited them. In this way, we kept our selected works quite up-to-date, including the most recent works with respect to this work. Figure 2 shows the distribution of the final selected works in terms of their publication year. As can be seen, more than half of the publications have been published recently (56% after 2014). Finally, the selected publications consist of three types, namely peer-reviewed journals, international conferences and workshops, as well as white papers in acknowledged academic repositories and archives. Figure 3 shows the distribution according to the publication type.

The remainder of the paper is organized as follows. In the next section, we present aspects regarding the definition and the effect of influence in user generated content. In Section 3 we analyze why semantics are significant for getting valuable and tangible insights for the disseminated information among users and social communities. Then, in Section 4, we highlight the qualitative assessments and positive impact of modeling the dynamic properties of influential content through semantics in OSNs. Finally, in Section 5 we present the conclusions of the surveyed works with respect to the reviewed topics along with our contributions in the domain.

[1] https://scholar.google.com
[2] https://arxiv.org/
[3] https://www.ssrn.com/
[4] http://dl.acm.org/
[5] http://ieeexplore.ieee.org/Xplore/home.jsp
[6] https://www.elsevier.com/catalog?producttype=journals

**Table 1:** Indicative keywords used to search appropriate publications (many were used combined with the “AND” Boolean operator in conjunction with terms such as OSNs, online social networks, social media, and so on, as a case insensitive search)

| | | |
|---|---|---|
| Influence | Social semantic modeling | Tweet quality |
| Influence maximization | Context-dependent influence | Event detection |
| Influence propagation | Content-driven approach | Information quality |
| Information propagation | Diffusion | Query expansion |
| Social network semantics | Sentiment-based influence | Social information retrieval |
| User interest | Similarity | Social recommendation |

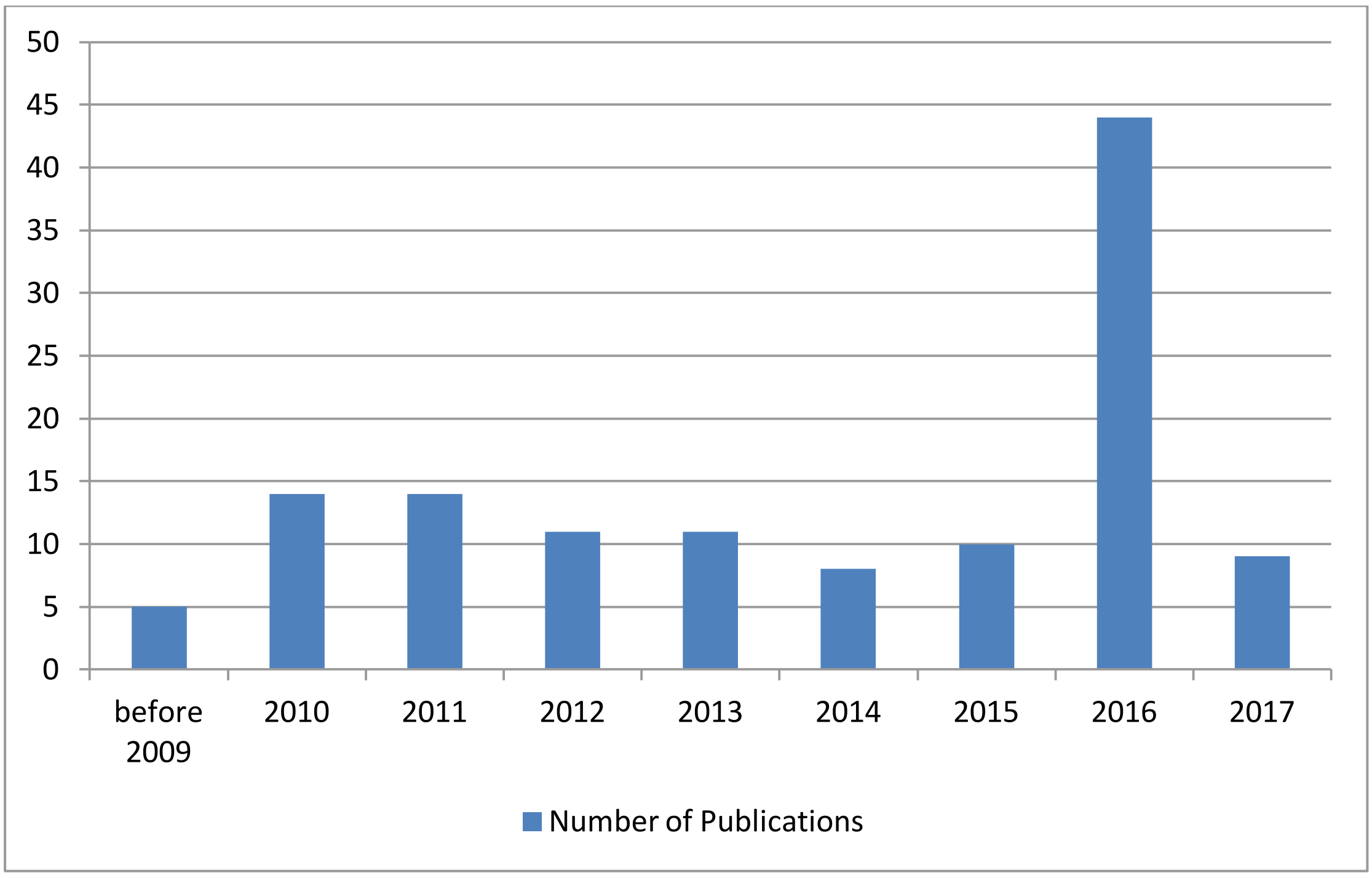

**Figure 2:** Number of publications per year for the selected works using IEEE Xplore, ACM Digital Library, Elsevier, Google Scholar, arXiv, SSRN

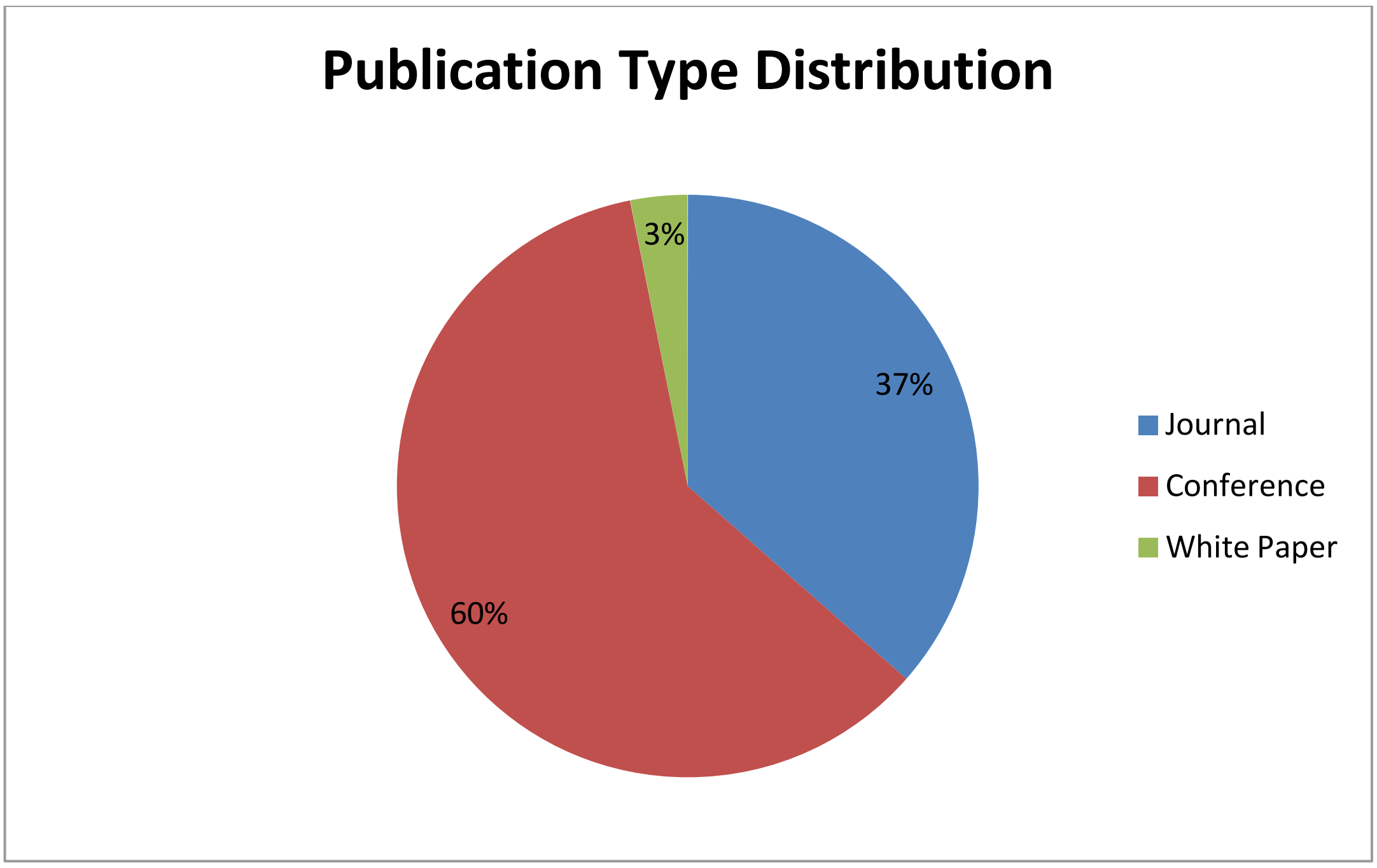

**Figure 3:** Distribution of publication type of all selected works

# 2. Related Work

In this section, we present other survey papers from the related literature that tackle similar issues in terms of modeling online social influence with semantics. Then, we highlight the differentiation and the added value of this review, as well as our contributions across our classification scheme.

## 2.1 Similar Surveys

As already mentioned, the reviewed works were classified into the hierarchical scheme depicted in Figure 1, resulting in 20 hierarchically structured categories. For the purposes of this extensive review, we also considered other survey papers that tackle similar aspects in terms of the impact of influence in OSNs and the role of semantics ([15], [85], [98], [106] and [107]).

More specifically, the authors in [15] focused mainly on the classification of current diverse measurements aimed at discovering influential users in Twitter. Their range varies from those based on simple metrics provided by the Twitter API to the adoption of PageRank algorithm and its variations. Other important factors are the content of the messages as some are focused on specific topics, their quality, in terms of likeability by others, as well as the activity and popularity of the users. Thus, the authors of [15] covered four aspects of our suggested scheme, namely “Influence Metrics”, “Network/Graph Properties”, “Social Matching: Topic and Event-oriented”, and “Qualitative Assessment”.

In [106], the authors analyzed a variety of OSN-based measurements and examined factors capable of affecting user influence. Those metrics were grouped under various criteria derived from:

- Neighborhood attributes, including number of influencers, exposure to direct and indirect influence.
- Structural diversity metrics that quantify the activity of the communities.
- Influence of locality and decay.
- Temporal measures including time delay until the reposting of a message.
- Cascade-based criteria, including its size and path length of messages.
- Metadata existence, including the presence of links, mentions, or hashtags.

Moreover, experiments were performed to predict user influence by using machine learning algorithms, with the aforementioned measurements as features. Based on our classification of this work, the survey described in [106] covers the "Applications: Ranking" and "Network/Graph Properties" categories.

The work in [85] presents an overview of studies regarding Adaptive Seeding (AS) methodologies to solve the Influence Maximization (IM) problem. IM is the process of discovering and activating a set of seed influential nodes-users to initiate the diffusion process so that the largest number of nodes is reached or influenced. Often, that set of users is restricted to those who are engaged with the topic of interest, and due to structural dependencies of the network, it is possible to rank low in terms of their influence potential. An alternative approach is to consider an adaptive method which aims at seeding neighboring nodes of high influence. As both IM and AS methodologies include the activation of nodes which in turn propagate the received information and activate others, the work described in [85] covers only the "Information Flow and Influence: Propagation-oriented" category as described in our survey.

The authors in [98] have reviewed approaches that enable Information Retrieval (IR) tasks in OSNs, which exploit content and structural social information. The research works the authors have reviewed have been classified into three categories according to the use of social information. Specifically, the "social web search" category includes techniques where social content is used to improve classic IR processes such as re-ranking of retrieved documents, query reformulation, expansion or reduction, and user profiling. The second category, called "social search", includes methodologies on information discovery based on users' generated content, interactions, and relationships. Finally, "social recommendation" aims at predicting users' interests and is based on content-based and collaborative filtering approaches. Hence, the survey in [98] covers the aspects of "Social Matching: User-oriented", "Network/Graph Properties", and "Applications: Recommendation", as described by this work.

Finally, the authors of [107] provide an overview on various user classification methodologies in OSNs. More specifically, they describe the most common frameworks based on machine (i.e. Bayesian, Decision Tree, Logistics, SVM and KNN) and non-machine (concept of entropy and based on user similarity) learning techniques. The aim of these methodologies is to classify users into certain categories according to their explicit or implicit features, such as behavioral attributes, profile information, interests, viral content and interactivity. As a result it covers only the "Social Matching: User-oriented" category as we have presented in this survey.

## 2.2 Review Differentiation and Extension

Table 2 provides comparative insights of this work in respect to the surveys described in [15], [85], [98], [106] and [107].. It consists of three columns. For every single survey, the first two represent the category and sub-category according to our classification scheme (Figure 1), as well as the respective section where we analyze it. The third column depicts the respective reference number. The mark "×" is placed in case where –according to the best of our knowledge- there is no other similar survey that covers this category.

Table 2. Classification of referenced surveys

| Category / Subcategory | | Section | Ref |
|---|---|---|---|
| Influence Metrics | Direct social information-based | 3.1.1 | 15 |
| | Hyperlink-based | 3.1.2 | 15 |
| | Machine Learning techniques-based | 3.1.3 | 15 |
| Information Flow and Influence | Propagation-oriented Approaches | 3.2.1 | 85 |
| | Diffusion-oriented Approaches | 3.2.2 | × |
| Network / Graph Properties | | 3.3 | 106 |
| | | | 15 |
| | | | 98 |
| Applications | Ranking | 3.4.1 | 106 |
| | Recommendation | 3.4.2 | 98 |
| | Other Application Domains | 3.4.3 | × |
| Social Modeling | | 4.1 | × |
| Social Matching | User-oriented | 4.2.1 | 107 |
| | | | 98 |
| | Topic and Event-oriented | 4.2.2 | 15 |
| Community Detection | | 4.3 | × |
| Qualitative Assessment | | 5 | 15 |

Thus, comparing to the surveys presented in Section 2.1, this review aims at covering and analyzing four additional aspects of OSNs, namely ”Information Flow and Influence” (further categorized in “Propagation-oriented” and “Diffusion-oriented”), “Social Modeling”, “Community Detection”, as well as ”Other Application Domains” related to online influence (Table 2). Moreover, the differentiation and extra issues covered in this work can be summarized in the following points:

- Information Flow and Influence: In contrast to many research works where the terms “diffusion” and “propagation” are used interchangeably, we tried to explicitly differentiate them by providing a clear distinction between their impact and role in the disseminated information.
- Social Modeling: we include studies aiming at the transformation of unstructured social data into Linked Data, by i) relating entities to knowledge bases (e.g. Google Knowledge Graph, DBpedia), and ii) representing them as concepts extracted from ontologies using semantic vocabularies.
- Community Detection: we consider approaches that also employ social semantics and ontologies. Such approaches are not only useful for the analysis of OSNs, but also for understanding the structure and properties of complex networks.
- Other Application Domains: Our survey also considers additionally topics and approaches, exploiting social influence for analyzing sentiment and user polarity, as well as detection of critical real life events.

## 2.3 Our contributions

In this work, we also place our approaches and proposals across the review classification scheme, and moreover in respect to:

- *Influence Metrics*: In [111] and [112], we present a publicly available service[7] aiming at calculating and ranking the importance and influence of Twitter accounts. Specifically, we define “*Influence Metric*”, which value derives from a social function incorporating i) the activity of a Twitter account (e.g. tweets, re-tweets, replies, mentions), ii) its social degree (e.g. followers, following) and iii) its impact in Twitter (e.g. content diffusion, social acknowledgement etc.).

- *Information Flow and Influence: Propagation-oriented and Diffusion-oriented*: In [110] we examined the propagation features and patterns of the Reddit social network. Specifically, the study is focused on the virality, lifespan, flows of information, and the speed of diffusion of user generated content across other OSNs.

- *Qualitative Assessment*: By extending the work described in [111], we introduced a new qualitative factor based on the established *h*-index metric ([112]). Its aim is to reflect other users’ actions (e.g. retweets) and preferences

[7] http://www.influencetracker.com/

(e.g. favorites, mentions) over the content and properties of viral posts, thus enhancing the "*Influence Metric*" of real influencers in Twitter.

- *Social Modeling*: In [112] and [114] we presented an ontological schema towards the semantification of social analytics. Specifically, the *"InfluenceTracker Ontology"*[8] is capable of modeling structural aspects of Twitter accounts, including information of their owners, all of their disseminated entities (mentions, replies, hashtags, photos, and URLs), as well as their online social relationships. In order to provide a five-star data model, according to Tim Berners-Lee's Linked Open Data (LOD) rating system [124], in [114] and [117] we extended our ontological schema by incorporating properties from DBpedia, foaf[9] and yago[10] ontologies. Since the latest update of the LOD[11] cloud, on 20/02/2017, the *InfluenceTracker* dataset is officially part of this interlinked and interdependent ecosystem of data. Finally, in [117] we propose an approach towards the automatic labeling of Twitter accounts in respect to DBpedia thematic categories.
- *Social Matching:* In [113], we proposed a methodology towards the discovery and suggestion of similar Twitter accounts. We consider term matching of user-generated content for all possible Twitter entities that may be used (mentions, replies, hashtags, URLs) according to [112] and [114]. Finally, we contributed in the field of users' query expansion in [115] and [116]. Specifically, we propose an algorithmic approach, which expands a user's query by creating a suggestion set of the most viral and up-to-date Twitter entities (e.g. hashtags, user mentions, URLs, etc.).

# 3. Online Social Influence

In this section we describe one major aspect in OSNs, namely, online social influence. We focus on how influence can be measured or predicted and the methodologies (e.g. based on topology, diffusion or social authority) that can be used to measure influence along with the respective application domains.

## 3.1 Influence Metric

The calculation of the impact a user has on social networks, as well as the discovery of influencers on them is not a new topic. It covers a wide range of sciences, ranging from sociology to viral marketing and from oral interactions to OSNs. In the related literature there is no strong agreement on what is meant by the term "influential user". Therefore, the term "influence" has multiple interpretations. Consequently, emerging influence measures are constantly varying with each of them using different criteria. Despite this variation, all the related studies do share a common result, which is that the most active users or those having the most followers are not necessarily the most influential ones. The works presented in this section discuss issues related to the discovery of influence and we classify these issues into three categories according to the way they a) exploit direct social information (number of followers, followees, social content, and so on), b) incorporate PageRank and related hyperlink-based algorithms, and c) employ machine learning techniques.

[8] http://www.influencetracker.com/ontology
[9] http://xmlns.com/foaf/spec/
[10] https://www.mpi-inf.mpg.de/departments/databases-and-information-systems/research/yago-naga/yago/#c10444
[11] http://lod-cloud.net/versions/2017-02-20/lod.svg

### 3.1.1 Direct Social Information Metrics

The study in [2] proposes the "Social Networking Potential" as a quantitative measurement for discovering influential users in Twitter, and suggests that having a large number of followers does not guaranty high influence. The methodology is based on the number of tweets, replies, retweets, and mentions of an account.

The authors in [12] introduce three types of influence, namely "In-degree" (number of followers), "Retweet" (number of user generated tweets that have been retweeted) and "Mention" influence (number of times the user is mentioned in other users' tweets) for Twitter users. A necessary condition for the computation of these influence types is the existence of at least ten tweets per user. The authors claim that "Retweet" and "Mention" influence correlate well with each other, while the "In-degree" does not. Therefore they conclude that the most followed users are not necessarily influential

Influence in terms of activity or passivity for Twitter users is studied in [22]. To conduct this study, a large number of tweets are utilized containing at least one URL, their creators and their followers. The influence metric produced depends on the "Follower-Following" relations of the users, as well as their retweeting behavior. As most studies in this area, it is stated that the number of followers a user has is a relatively weak predictor of the maximum number of views a URL can achieve.

Another metric for calculating the importance and influence of a Twitter account called "Influence Metric" is proposed in [111]. In this case, a social function is presented which incorporates the activity (or passivity) of a user, and the number of followers and followees. Moreover, the same authors in [112] introduce an improved version of the aforementioned metric, by incorporating a new qualitative factor according to the established *h*-index factor whose aim is to reflect other users' actions and preferences over the content of the created posts, thus enhancing the influence of quality users. H-index is also used in [22] as a predictor of high achievements for retweets containing URLs.

In [3] the "*t*-index" metric is proposed, which aims at measuring the influence of a user on a specific topic. It is also based on the *h*-index factor and denotes the number of times a user's tweet on a specific topic has been retweeted. The authors suggest that high influence on one topic does not necessarily mean the same on other topics.

A framework exploiting influence for evaluating and enhancing communication issues between governmental agencies and citizens in OSNs is proposed in [118]. The aim here is to evaluate the quality of the agencies' responses in respect to the citizens' requests, to analyze the citizens' sentimental attitude and their subsequent behaviors, and to suggest influential users to the agencies in order to obtain new audience. To achieve these, several components are incorporated into the framework, which detect the demographics of the followers, their locations, topics of interest, and sentiments.

The authors in [91] propose a different kind of influence called the "susceptibility to influence". Its metric estimates how easily a Twitter user can get influenced. The proposed metric utilizes the user's social interactions that depend on three factors namely, activity, sociability and retweeting habit. Activity reflects the user's tendency to interact with friends and consequently the chance to become influenced by them, while "sociability" corresponds to the users' social degree among their activities, implying that interactions with more friends result in a wider diversity of topics and interests.

Finally, the study in [97] presents a methodology for measuring social influence in mobile networks by incorporating its entropy. Specifically, the friend and interaction frequency entropies are introduced in order to describe the complexity and uncertainty of social influence. A weighted network is constructed based on users' interactions, upon which three types of influence are introduced, namely a) direct influence among related users, b) indirect influence among unrelated users, and c) global influence that covers the whole network.

### 3.1.2 Hyperlink-based Metrics

The studies in this sub-section describe influence metrics through hyperlink-based algorithms (e.g. PageRank), and therefore influence is strongly related to the structure and the topology created by the OSN itself.

The authors in [77] propose an influence model based on two aspects, user relationship and activity. They used three factors namely Influence Diffusion Model (IDM), PageRank, and usage behavior. IDM focuses on tweets and their reply chain, while providing the influence of propagation based on word occurrence. The PageRank algorithm is employed for calculating users' significance based on their relationships, while the user behavior factor affects a user's influence score as it is based on the number of posts, mentions, follower, and retweets. The core ideas of these models are extracted and are integrated into the proposed influence model.

An influence ranking method is proposed in [19] based on the fact that the influence of a user is determined by the followers' influence contribution, which in turn highly depends on their interactions. A user can exert more influence over another if the former writes more tweets related to those of the second user. The proposed measurement is a variation of the PageRank algorithm and is based on a similarity factor between published tweets over a graph of following users, retweets, mentions, and replies.

In [18], the proposed "Influence Rank" metric implements a modified version of the PageRank algorithm, which is based on the structure and topology of the network. Specifically, it combines follow-up relationships, mentions, favorites and retweets to identify opinion leaders who are capable of influencing others. The authors conclude that in order to be influential, a user should have influential followers.

In [21], the authors present a variation of the PageRank for introducing two metrics called "InfRank" and "LeadRank", which are based on following, retweeting and mentioning relationships among users. "InfRank" is a variation of PageRank and measures the user influence in terms of his/her ability to spread information and to be retweeted by other influential users. "LeadRank" measures the leadership of a user in terms of his/her ability to stimulate retweets and mentions from other users and especially from other leaders.

Finally, in [123], the authors present the "MISNIS" framework whose goal is to discover influential Twitter users on a given topic. The framework does so by applying the PageRank algorithm on a graph representing users' mentions found in Portuguese tweets. Moreover sentiment analysis is performed, classifying the messages into three categories namely, positive, neutral, and negative. This work differentiates itself from others in this field in the way that the topics are detected. Instead of performing naive string matching based on the characters of a hashtag, a fuzzy word similarity algorithm is applied utilizing all the contents of a message. Consequently, more relevant tweets for a topic are retrieved despite not containing the exact hashtags or other user indicated keywords.

### 3.1.3 Metrics based on Machine-learning Techniques

In [87], social influence is measured by applying the "InfluenceRank" framework. It is based on certain features extracted from profiles (number of tweets, followers, following, member of lists) and tweets over a two-month period. The framework comprises of a regression-based machine learning approach, having "InfluenceRank" as the predictor variable against the set of aforementioned features. Although the work seems promising, the authors claim that, due to the limited number of samples in the training set, the model is not accurate enough.

Another machine learning framework for discovering popular persuasive users is presented in [99]. The authors' persuasiveness metric is pair-wise and is based on three factors: influence, entity similarity, and structural equivalence. Influence depends on the strength of social interactions among users, entity similarity measures how close two profiles are, while structural equivalence measures the structural similarity of two entities according to a distance function. Each of these factors is assigned a probability which denotes the likelihood of persuasion.

The work presented in [7] proposes a framework for predicting user influence by combining textual and non-textual attributes. More specifically, the authors employ the user's basic social information metrics (e.g. the number of followers, followees, mentions and replies), and then by utilizing statistics over the textual data of the tweets, as well as non-linear learning methods and machine learning techniques, a strong prediction performance metric is derived.

Finally, the authors in [122] propose a machine learning methodology for investigating the impact of profile information towards the increase of Twitter accounts' popularity, in terms of their followers' count. Based on the assumption that given names and English words affect the discoverability, profiles were analyzed and categorized into three groups according to the lexical content of the accounts' name: i) having a first name, ii) containing English words, or iii) neither of both. The framework consists of three stages to evaluate the popularity dynamics in terms of: a) the content of the name field, b) the profile features, and c) the incorporation of those features in a classifier that identifies the accounts which are likely to increase their popularity. Each group's classifier uses a different model (e.g. Gradient Boosting Machine, Naive Bayes), which is trained with distinct parameters and features, based on the corresponding group. The results showed that the existence of known terms in the name field and the provision of other profile information (e.g. description, profile image, URLs, location) have a strong impact on the number of followers.

## 3.2 Information Flow and Influence

Information flow is vital in all kinds of networks (e.g. social, digital, or computer), and can be affected by the actions or properties of their actors and the sets of dyadic relationships between them. Influential users determine the virality of information and specifically how such information is propagated or diffused. As already mentioned, although propagation and diffusion are often used interchangeably, in this survey we examine them separately. Diffusion defines the spread of information from a starting node towards the rest of the network, while propagation takes into consideration the intermediate nodes as well, which receive, process, and further decide how to handle information. Thus, this topic is divided into two subtopics namely, propagation-oriented and diffusion-oriented. The propagation-oriented approach considers works that employ the propagation of information in OSNs in order to discover and calculate the impact of influential users whereas the diffusion-oriented approach provides the insights into the identification of influential users being able to boost the diffusion of information in OSNs.

### 3.2.1 Propagation-oriented Approaches

The authors in [17] propose an extension of PageRank for measuring influence. They applied their extended PageRank approach on a graph of retweets and user relationships and consider social diversity of users and transmission probabilities of the messages based on the hypothesis that users inherit influence from their followers. The aim is to explore whether individual characteristics and social actions as well as influence propagation patterns are factors capable of influencing other users.

Similarly, as described in the previous subsection, in [21] social influence is measured using a variation of PageRank. Specifically, the authors measured the propagation of user influence into the network based on the users' ability to stimulate social actions of others, such as retweets and mentions.

In several cases, the influence metric derived correlates the information propagation with the user's retweeting behavior. Such a study is described in [22], where influence is used for measuring the activity or passivity of Twitter users.

The authors in [76] propose a methodology to identify influencers in OSNs with the help of online communities which are discovered by applying propagation-based modes. In this case, the structural features (shortest path, closeness, eccentricity, betweeness, and degree) of each node are extracted, while their weighted representation is computed by considering all the features across the network. By using principal component analysis, the most influential nodes are discovered. By applying maximum flow algorithms communities are detected implying positive attitude towards the influencers.

Social influence and propagation can be used as input in recommendation systems. In [14], influence is considered as a propagated attribute among users in OSNs. The proposed framework calculates the influence that social relationships have on users' rating behaviors, and incorporates it into recommendation proposals. Two social influence related attributes are considered: user's susceptibility, which is the willingness to be influenced, and friends with high influence.

While the above studies consider the propagation of information towards the discovery of influential users, there are many other works ([27], [42], [75], and [95]) that describe frameworks for discovering the propagation of influence in Twitter and its impact on other users.

A framework for modeling the spread of influence in OSNs is developed in [27]. The authors characterize influential users as those generating posts with high probability of being propagated, i.e. retweeted, and simultaneously having a large number of followers. Based on past information cascades influential users are discovered and their social activities and interconnections inside the communities they belong to are analyzed.

In several works, the influence of a node is calculated based on the rate of information spread over the network. For each influenced node an influence function quantifies how many subsequent ones can be affected. This is based on the assumption that the number of newly influenced nodes depends on which other nodes were influenced before. The study described in [42] concludes that the diffusion of information is governed by the influence of individual nodes. Similar to the previous study [27], the proposed models are considered as stochastic processes in which information propagates from a node to its neighbors according to a probabilistic rule. The problem lies in discovering influential nodes based on the computation of the expected number of influenced ones [43].

A study analyzing the persuasion-driven social influence based on some topic of interest is presented in [95]. Several influence measurements incorporate users' social persuasiveness in terms of influence propagation, for quantifying user-to-user influence probability. Based on proposed metrics, the framework exploits the topical information, the users' authority and the characteristics of relationships between individuals.

A multi-topic influence propagation model is proposed in [75]. It is based on user relationships, posts, and social actions. The influence score consists of direct and indirect influence, where the former considers information propagation from retweets by the direct followers, while the latter takes into account the retweets from non-followers. Both of them are related to different topics. The distribution of users' topics of interest is discovered according to the collected tweets. Then, a topic-dependent algorithm is applied and a multi-topical network is created, in order to identify multi-topic influential users.

A model for demonstrating how social influence can impact the evolution of OSNs by simulating influence propagation and activation processes is proposed in [90]. In this model, two types of influence namely, locality and popularity are considered since they have different impact on the network dynamics. Locality affects the information spread through social ties, while popularity has global impact on individuals since it does not rely on network topology.

Influence maximization (IM) problem is the process of discovering and activating a set of seed nodes to initiate the diffusion process so that the largest number of nodes is reached or influenced. The authors in [79] investigate the IM problem and propose a probability-based methodology that enables greedy algorithms to perform efficiently in large-scale social networks in terms of memory and computing costs. The algorithms recursively estimate the influence spread using reachable probabilities from node to node. In [96], the authors aim to maximize influence propagation by selecting the most influential intermediate nodes. Therefore, a new optimization problem is formulated which explores the idea of routing multi-hop social influence from the source to a specific target with some time constraint. To achieve this, the topology of the network, the users' influence and the responding probability in a specific time frame are taken into consideration. The authors in [92] propose a content-centered model of flow analysis in order to investigate the IM problem. Moreover, the analysis is not based on the users' relationships but on the content of the transmitted messages. The authors apply an algorithm to discover the information flow patterns using content propagation patterns. Then, the influencers are discovered by exploiting those patterns, their position and the number of flow paths they participate in. A different approach on the IM problem is proposed in [109] and is described as "boost set selection". The authors claim that it is possible to improve the diffusion process of a subset of the initial seed nodes by using additional resources such as by giving out free samples of a product, engaging in gamification, or other marketing strategies in order to become more influential.

An extension of the IM problem, described as "Influential Node Tracking", is defined in [88] where the authors focus on the set of influential nodes dynamically such that the influence spread is maximized at any time. Due to the dynamic nature of the networks, their structure and influence strength associated with the edges constantly change. Consequently, the seed set that maximizes the influence coverage should also be constantly updated. To achieve their goal, the authors compare consecutive snapshots of a network based on the fact that it is unlikely to have drastic changes thereby resulting in great structural similarity.

Finally, there are other works ([40], [44], and [110]) that investigated the discovery of information propagation flows in OSNs. In [40], a diffusion network is constructed based on user mentions, with constraints on topical similarities in the tweets. The authors claim that given the lack of explicit threading in Twitter, this is the optimal approach of a network to spread information about a specific topic, and that the rate of mentioning of a user is a strong predictor of information propagation. In addition, the authors in [44] examine information propagation that is related to the exposure to signals about friends' information sharing on Facebook. They found that the users who are aware of that information are significantly more likely to share it faster, compared to those who are not. Although these strong ties are individually more influential, the weak ties, which exceed them in numbers, are responsible for the propagation of information.

### 3.2.2 Diffusion-oriented Approaches

In [37], the authors investigated diffusion issues with an improved version of the K-core method [83]. The authors incorporate a linking and weighting method based on the observation that users' interactions, namely retweets and mentions, are significant factors for quantifying their spreading capability in a network. In [104] the authors propose the "SIRank" metric for measuring users' spread ability and identifying influential ones. Initially the users' spread influence is measured by analyzing the information cascade structure. As each user's influence is directly related to his/her interaction influence with others, pair-wise metrics are calculated by measuring retweeting contributions, users' interests and closeness, activity frequency, and retweeting intervals. By quantifying cascade structure and user interaction influence on information diffusion, the authors measure the users' spread influence.

Similarly, the main objective in [101] is to investigate the diffusion of messages and users' influence, based on the retweet cascade size and its attenuation patterns. The proposed influence measurement depends on the number of users who could potentially get a message either directly or via retweets. The latter affects the proposed cascade size metric and sets the upper limit of users who could potentially see that message. The study concludes that the largest cascades originate from users with most followers and, the cascade dies out after two or three frequency peaks.

The "retweet" functionality and the retweet counter can be considered as a factor for measuring the "interestingness" of a user's tweets [33]. Based on that, the resulting spread of information is examined in [11]. The authors state that the retweet counters are measurements of popularity of the messages and of their authors. According to the study, once a message gets retweeted, it will almost instantly be spread up to four hops away from the source, thus resulting in fast diffusion after the first retweet. Three different measures of influence namely, the number of followers, PageRank, and number of retweets, were further compared and evaluated. The results indicated that, in contrast to the third measurement, the first two provide similar rankings of influential users, indicating a gap in the influence derived from the number of followers and the popularity of the tweets. Similar to the results of [101], the average number of additional recipients is not affected by the number of followers of the tweet source. Thus the tweet is likely to reach a certain number of audiences via retweets.

A different interpretation of the term "influence" is given in [9], where the authors relate the user's posting activity (and thus influence) with the diffusion of the URLs included in posts through retweets. The influence score for a given URL post is calculated by tracking the diffusion of the URL from its source node until the diffusion event is terminated. The work is similar to the one described in [77] where the influence measurement is related with the Influence diffusion model which provides the influence of topical spread. However, it differs from [22] in that the diffused influence is studied in terms of activity or passivity of Twitter users solely based on the user's retweeting behavior.

In addition to the point of views discussed above, the following studies involve methodologies for analyzing information diffusion and factors that affect it in OSNs. As already described in the previous subsection, the authors in [40] claim that, despite the fact that some properties of the tweets predict great information propagation, the users' mention rate is the strongest predictor. The diffusion of information in two social networks, namely Digg and Twitter, is studied in [41]. According to the study, the structure of these networks affects the dynamics of information flow and spread. Information in denser and highly interconnected networks such as of Digg's, reaches nodes faster compared to sparser networks such as of Twitter's. Due to its structure, information is spread slower, but it continues spreading at the same rate as time passes and penetrates the network further.

In [44], information spread is examined regarding exposure to signals about friends' information sharing on Facebook. The study concludes that social ties greatly affect users' behavior on re-spreading information in the network. In another work on Facebook, the authors studied diffusion trees of fan pages. The results indicated that there is no solid evidence that a node's maximum diffusion chain length can be predicted [10].

The ways in which widely used hashtags spread through interactions among Twitter users are analyzed in [45]. Hashtags of different types and topics exhibit different variations of spread. The variations are due to the differences in the spread probability, and differences in the extent to which repeated exposures to hashtags continue to affect their diffusion into the network by other users. The authors in [108] extend their previous work [126] to identify the initial set of users who are able to maximize information diffusion. Initially, the users' diffusion patterns are recognized by exploiting their posting activities and history. The proposed algorithm combines them with propagation heuristics in order to achieve the diffusion coverage in the network.

Finally, the authors in [110] studied the lifespan and information flows of another social network (Reddit) based on user-generated content. They were particularly interested in the virality of information and its speed of diffusion in other OSNs.

## 3.3 Network/Graph Properties

The studies presented herein utilize the topology and the structure of the OSNs in order to measure influence or to discover other social dynamics. Usually, in this domain, only a fraction of the whole network is used due to hardware (e.g. RAM, Hard Disk Drive) or complexity limitations.

The framework proposed in [6] aims to automatically identify influential users in topic-based communities. Therefore, a sparser network of Twitter is created in comparison to the traditional follower/following network, by leveraging direct communications (mentions and replies). A measure of alpha centrality is employed, which incorporates both directionality of network connections and a measure of external importance. As already mentioned in [21] and [77], influencers are discovered by applying PageRank and newly proposed link-analysis algorithms which are exploiting the topology and properties of the network, including posting, retweeting and mentioning relationships among users. In [78] influence is measured by applying a hybrid framework that integrates both users' structural location and attributes. A user's location is found by applying several centrality analysis algorithms (in-degree, weighted in-degree, eigenvector, and PageRank) while the attributes (i.e. activeness) are measured by adapting the contribution measurement, which is used by Flickr, and is based on the number of uploaded photos.

The authors in [76] propose a methodology for the identification of influencers by exploiting structural features. Specifically, the shortest path, closeness, eccentricity, betweeness centrality, and degree of each node are extracted and their weighted representation is computed by considering all the features across the network. The most influential nodes are discovered by using principal component analysis. Moreover, by applying maximum flow algorithms communities are detected. The identified communities imply a positive attitude towards the influencers.

In [90], the authors proposed a framework to demonstrate how social influence can impact the evolution of OSNs by simulating influence propagation and activation processes. In this framework, two types of influence are introduced, that have different effects on the network dynamics. The first type is "locality" which affects information diffusion through social ties, while the second is "popularity" that does not rely on network topology but has global impact on individuals.

All above studies try to identify influencers according to the information derived in particular periods of time, similar to a compilation of different and static sequences. Below, we analyze other works where related issues are considered under properties and concepts that belong to dynamically evolved and complex networks.

Moreover, a dynamic index data structure for influence analysis on an evolving network is presented on [80]. The indexing method is able to recognize and incorporate all graph updates in order to efficiently answer the queries on influence estimation and maximization on the latest graph edition. Several optimized techniques (e.g. a reachability-tree-based technique for edge/vertex deletions, a skipping method for vertex additions, and a counter-based random number generator for the space efficiency) are incorporated to reduce time and space requirements.

In [88], the Influential Node Tracking problem is defined as an extension of the Influence Maximization one dynamically evolving networks. Due to their nature, the structure and influence strength associated with the edges change constantly. Therefore, the authors consider the dynamic network as a set of static ones, and compare consecutive snapshots under the assumption that it is unlikely to have drastic structural changes.

Another work in this area is presented in [100]. A dynamic network is modeled as a stream of edge weight updates. Under the assumptions of the linear threshold model, two versions of the problem are considered: the discovery of nodes having influence greater than a specified threshold, and finding the top-k most influential nodes. The proposed algorithm incrementally updates the sample random paths against network changes by considering efficiency in terms of both space and time usage.

Apart from discovering influential users, the topological and structural attributes of the networks can be used towards context-based identification of users' interests and similarities. For example, a community detection in OSNs approach is proposed in [50] using node similarity techniques. A virtual network is created, where virtual edges are inserted based on the similarity of the nodes in the original network. The similarity is calculated using the Jaccard Measure. The proposed algorithm is then applied on the generated virtual network.

Similarly, the author in [68] proposes a semantic followee recommender system in Twitter which exploits users' tweets in order to build their interest profiles. An interest graph is created by using specific semantic knowledge graphs that contain a variety of topics. These topics are then mapped and suggested to the users. User interest metrics are calculated using graph theory algorithms such as the Steiner Tree and the "InterSim" (Interest Similarity) one.

Another context-oriented approach is presented in [71], where the context of Twitter posts is retrieved using the DBpedia knowledge base and graph-based centrality theory. A graph of contextualized and weighted entities for each tweet is constructed, and two types of similarity metrics are introduced. The "local" similarity measures the proximity of two entities in terms of the context in which they occur. When a user request is made, the "global" similarity is calculated from this request and the available tweets.

## 3.4 Applications

In this section, we consider the influence metrics presented in Section 3.1, to present research efforts that provide solutions for opinion makers, data analysts and information scientists as services or applications. This topic is further divided in three subtopics namely a) rank-oriented, b) recommendation-oriented and c) other application domains, such as sentiment analysis, event detection, and so on.

### 3.4.1 Ranking

To rank OSN users according to specific social attributes, the work described in [5] presents a qualitative measurement of tweets that determines the influence of their authors in order to present a tweet-centric topic-specific author ranking. The quality of the tweet is evaluated according to the topic focus degree, the retweeting behavior, and the topic-specific influence of the users who retweeted it. In [19], the authors propose an influence ranking method under the assumption that the user influence is based on the followers' influence and their interactions. The authors found that user A can exert more influence over user B if user A posts tweets strongly related with user B. The proposed measurement is a variation of the PageRank and is based on a similarity factor between published tweets, on a graph of following, retweets, mentions and replies.

The authors in [70] propose a framework for discovering topic-specific experts in Twitter by employing two distinct metrics. First, the users' global authority (influence) on a given topic is calculated offline by exploiting three types of relations (i.e. follower relation, user-list relation, and list-list relation). Second, the similarity between the users' generated tweets and that topic is computed online. By leveraging the users' topical influence and similarity, those who have the highest-ranking scores are regarded as experts in that domain.

The problem of topic-sensitive opinion leaders' identification in online review communities is also investigated in [84], where a two-staged approach is presented. Initially, the opinion leaders' expertise and interests are derived from their tags found at the description of the products. Then, a computational approach measures the leaders' influence and ranks them according to not only the link structure of customer networks, but also according to their expertise and interests. The influence depends on the topical similarity between reviewers on a specific topic.

The authors in [86] created "NavigTweet", an influence-based visualization framework to explore Twitter's followers relationships, by browsing the friends' followers network and for identifying key influencers based on the actual influence of the disseminated content. The top influencers are identified by both user-level (i.e. followers, following, tweets, lists) and content-based (hashtags, URLs, retweets, favorites, mentions) parameters. Then, based on the above, a technique called "Analytical Hierarchy Process" ranks Twitter users.

An influence learning-based recommender is presented in [102] for making suggestions to informative users whose posts are highly associated with those of target users. Ranking learning techniques are designed to analyze user behavior and to model their preferences based on their social interactions (e.g. replies, likes). Moreover, the social influence among users is incorporated into the learning model to enhance the learned preferences. In another application described also in Section 3.2.2 the authors propose the "SIRank" metric for measuring users' spread ability and identifying the influential ones [104].

Finally, "InfluenceTracker" is a publicly available service[12] capable of calculating and ranking the importance and influence of a Twitter account [111] and [112]. The authors introduce a social function, which incorporates the activity (or passivity) of a user, his/her popularity (e.g. number of followers and followees), and his/her social acknowledgment by others users.

### 3.4.2 Recommendation

The various studies presented in this section describe approaches on recommendation systems which utilize the available information in OSNs for proposing social content or accounts based on the users' profiles. An interesting problem in the area of social network recommendation systems is to define a set of similar users to follow.

The friend recommendation problem in Flickr is studied in [63], mainly from the viewpoint of network correlation. The authors assume the hypothesis that each user has many different social roles in OSNs. For each role different social sub-networks are formed, which are aligned in order for the correlations among them to be found through weighted tag feature selection. When recommendations are made, the similarities of the tag features, among the new and the existing users, are calculated. The more similar the tags are, the more users we have who are similar in terms of those tags.

A semantic followee recommender system in Twitter is proposed in [65]. This system integrates content-based filtering approaches based on tweet analytics, popularity identification among users using collaborative-filtering over the friendship network, along with publicly available knowledge resources (i.e. Wikipedia, WordNet, Google corpus). The aim is to classify the tweets into six classes and to label the users as a recommendation service. The application of the Kalman filter enables noise removal and the prediction of future tweet patterns leading to the new multi-labeling of the users.

Similarly, the work described in [68] exploits the users' tweets for building their interest profiles and for producing recommendation over a semantic knowledge graph that contains a variety of topics. Using graph theory algorithms (as explained in Section 3.3), the authors can recommend similar users. A ranking-based followee recommendation scheme in microblogging systems that is based on latent factor model is proposed in [73]. To model user preferences both tweet content (original posts and retweets) and social relation information (followers, followees) are taken into consideration. Another followee recommendation methodology that builds interest profiles is proposed in [35]. These profiles are built by exploiting not only the users' generated content but also of their directly related ones (followers, followees).

A framework for discovering similar accounts in Twitter based only on the "List" feature is proposed in [64]. This functionality allows the users to create their own lists by adding any account they wish. The authors claim that this feature is considered a form of crowdsourcing. The hypothesis of the methodology is that when two accounts are contained in the same list they should be similar or related to each other. Therefore, the proposed measurement relies on the number of lists that a specified account and a potentially similar one are listed together.

[12] http://www.influencetracker.com/

In [38] a matrix factorization framework with social regularization is proposed for improving recommender systems by incorporating social network information. Social regularization includes two models for representing social constraints and are based on users-friends similarity at an individual and average level. Each social link is then weighted in accordance with the similarity among the users, allowing the exploitation of friends based on the rating similarity.

As already described in the previous subsection, ranking learning techniques are designed to provide recommendation based on the analysis of user behavior, preferences, and social interactions [102]. In addition, as mentioned earlier in Section 3.2.1, we can use related attributes being propagated through the social network because the effects of friends who have strong influence or are subject to be influenced by a user, are highly related with recommendation processes [14].

The recommendation system proposed in [103] is based on the users' personal interests. In fact, explicit social features such as the users' topic-level influence, topic information, and relations are incorporated into a framework for improving recommendation results. Two kinds of influence are introduced: direct, which is identified by studying the communication records between users, and indirect, which is identified by applying social status theory for the discovery of latent relationships. In both cases positive and negative influences are also identified. Moreover, topic information is added into the structural analysis of indirect influence. A distributed learning supervised algorithm is applied which takes into consideration the aforementioned influence measurements and provides the users' forwarding behaviors, which can be leveraged to provide improved recommendations.

A lot of attention has also been paid to recommendation systems for suggesting personalized streams of information ([34], [36], [75], and [82]).

"Buzzer" is such kind of a service for proposing news articles to Twitter users, by not only mining terms from their timeline, but also from their friends [34]. These terms act as ratings for promoting and filtering news content. The methodology described in [66] is based on the same principles but it also incorporates additional factors affecting the interest of a user on a tweet, such as its quality, number of retweets, and the importance of its publisher.

URLs as a recommendation factor in Twitter are studied in [36] in terms of directing users' attention in more focused information streams, namely Twitter posts, from the viewpoint of personalized content suggestion. The authors explored three separate dimensions in designing such a recommender: the sources of the URLs, the users' area of interest, and social information.

The authors in [75] propose a multi-topic influence diffusion model based on user relationships, posts, and social actions. The influence score consists of direct and indirect influence. The first is determined by information propagation (retweets) by the direct followers. The latter depends on the retweets from non-followers. Both of them are related to different topics. Based on the users' collected tweets, their distributions of topics of interest are found along with their generation probability. Finally, a multi-topical network is created to which a topic-dependent algorithm is applied in order to identify the multi-topic influential users while the most influential user will be used during the recommendation process.

Finally, recommenders can also be used for suggesting items on users. The work described in [82] is based on the observation that user's purchase behavior is influenced by both global and local influential nodes which in turn define implicit and explicit social relationships respectively. Therefore, a dual social influence framework formulates the global and local influence scores as regularization terms, and incorporates them into a matrix factorization-based recommendation model.

### 3.4.3 Other Application Domains

The studies presented in this section exploit users' influence in OSNs in other domains (such as sentiment analysis and user polarity as well as event detection) than those described in Sections 3.4.1 and 3.4.2.

Moreover, the identification of influential users in Twitter is based on a combination of the users' position in networks derived from Twitter relations, the sentiment of their opinions, and the textual quality of their tweets. Thus, in [8], the authors propose a centrality measure that combines betweeness and eigenvector centralities, in-degree and the follower-followee ratio on graphs of relationships, mentions, replies and retweets. Using sentiment analysis techniques, the users are classified into those having positive, negative or neutral tweets.

Another sentiment-based framework is proposed in [28], where sentiment is discovered through exchanged messages among users in online health communities. The metric focuses on the sentimental effect of inter-personal influence on individuals and reflects a user's ability to directly influence other users' sentiments.

The authors in [93] investigate whether the users' friendship network can interfere with the peer and external influence. The experiment takes place during an on-line voting procedure in Facebook. The analysis of the users' demographics and votes showed strong homophily among the communities and friends' votes. The authors analyzed both peer and external influence in order to explain the activation of voters. Peer influence propagates from recently activated friends while external influence from news agencies affects all users uniformly.

Finally, a story-tracking framework based on hashtags in OSNs is proposed in [72]. The storyline extraction is modeled as a pattern mining and real-time retrieval problem. The most popular news stories, which have been assigned hashtags, are detected by mining frequent hashtag pattern sets. Using query expansion on the original hashtags new story articles are retrieved. The pattern set structure enables hierarchical and multiple-linkage representation of the news.

## 3.5 Comparison of Related Works

In order to provide comparative insights from the above reviewed articles that refer to online social influence, we provide Table 3. For each reviewed article, the first three columns denote its category according to our classification scheme (see Figure 4 below), the section number, as well as its reference number. It should be noted that in many cases, a study does not fall with the scope of only one topic, thus demonstrating that research efforts are strongly related to each other. In the rest columns, we place a mark of “Yes” (✓) or “No” (✗), in case where the studies employ or propose metrics and characteristics based on:

a. Relationship: followers (Fs) and followees (Fing),

b. Behavioral/Conversational activities: posts (P), re-posts (RP), favorites/likes (FL), mentions (M), replies (R), and

c. Domain/Content analysis: works that are applied on specific topics (T) or works that require content analysis (CA).

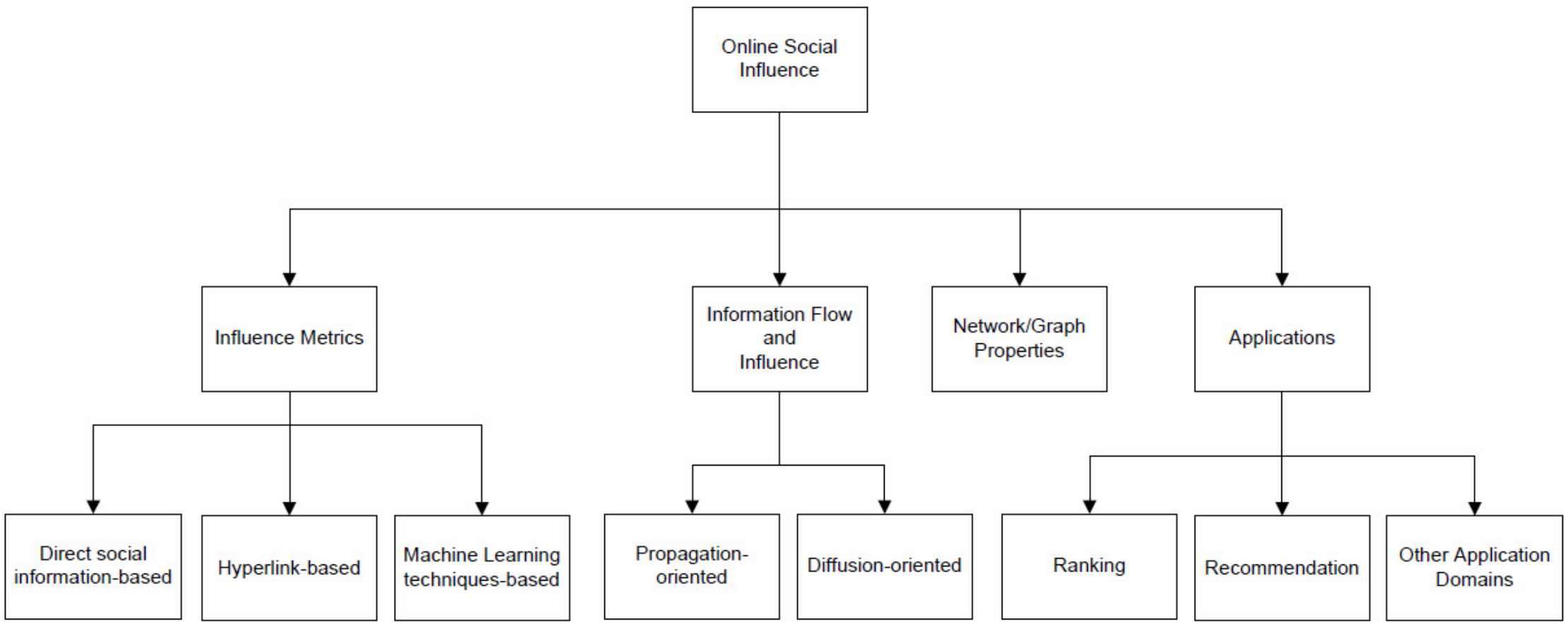

Figure 4: The hierarchical classification scheme followed in this work on Online Social Influence [part of Figure 1]

Table 3: Classification of referenced works. Fs and Fing refer to follow-up relationships, P to posts, RP to reposts, FL to favorite or liked posts, M to mentions, R to replies, T to topics, and CA to content analysis.

| **Category** | **Section** | **Ref** | **Relations** | | **Activities** | | | | | **Context** | |
|---|---|---|---|---|---|---|---|---|---|---|---|
| | | | **Fs** | **Fing** | **P** | **RP** | **FL** | **M** | **R** | **T** | **CA** |
| Direct social information-based | 3.1.1 | 2 | ✓ | ✗ | ✓ | ✓ | ✗ | ✓ | ✓ | ✗ | ✗ |
| | | 3 | ✗ | ✗ | ✗ | ✓ | ✗ | ✗ | ✗ | ✗ | ✗ |
| | | 12 | ✓ | ✗ | ✓ | ✓ | ✗ | ✓ | ✗ | ✓ | ✗ |
| | | 22 | ✓ | ✗ | ✓ | ✓ | ✗ | ✗ | ✗ | ✗ | ✗ |
| | | 91 | ✗ | ✗ | ✓ | ✓ | ✓ | ✓ | ✓ | ✓ | ✗ |
| | | 97 | ✗ | ✗ | ✗ | ✗ | ✗ | ✗ | ✓ | ✗ | ✗ |
| | | 111 | ✓ | ✓ | ✓ | ✓ | ✓ | ✗ | ✗ | ✗ | ✗ |
| | | 112 | ✓ | ✓ | ✓ | ✓ | ✓ | ✗ | ✗ | ✗ | ✗ |
| | | 118 | ✓ | ✗ | ✓ | ✗ | ✗ | ✓ | ✓ | ✗ | ✓ |
| Hyperlink-based | 3.1.2 | 18 | ✓ | ✓ | ✗ | ✓ | ✓ | ✓ | ✗ | ✗ | ✗ |
| | | 19 | ✓ | ✓ | ✓ | ✓ | ✗ | ✓ | ✗ | ✗ | ✓ |
| | | 21 | ✓ | ✗ | ✗ | ✓ | ✗ | ✓ | ✗ | ✗ | ✗ |
| | | 77 | ✓ | ✓ | ✓ | ✓ | ✗ | ✗ | ✓ | ✗ | ✓ |
| | | 123 | ✗ | ✗ | ✗ | ✗ | ✗ | ✓ | ✗ | ✓ | ✓ |
| Machine Learning techniques-based | 3.1.3 | 7 | ✓ | ✓ | ✓ | ✗ | ✗ | ✓ | ✓ | ✗ | ✗ |
| | | 87 | ✓ | ✓ | ✓ | ✗ | ✗ | ✗ | ✗ | ✗ | ✗ |
| | | 99 | ✗ | ✗ | ✗ | ✗ | ✗ | ✗ | ✗ | ✗ | ✗ |
| | | 122 | ✗ | ✗ | ✗ | ✗ | ✗ | ✗ | ✗ | ✗ | ✓ |

| Category | Section | Ref | Relations | | Activities | | | | | Context | |
|---|---|---|---|---|---|---|---|---|---|---|---|
| | | | Fs | Fing | P | RP | FL | M | R | T | CA |
| Propagation-oriented Approaches | 3.2.1 | 14 | ✓ | ✓ | ✓ | × | × | × | × | × | × |
| | | 17 | ✓ | × | × | ✓ | × | × | × | × | × |
| | | 21 | ✓ | × | × | ✓ | × | ✓ | × | × | × |
| | | 22 | ✓ | × | ✓ | ✓ | × | × | × | × | × |
| | | 27 | ✓ | × | ✓ | ✓ | × | × | × | × | × |
| | | 40 | × | × | × | ✓ | × | ✓ | ✓ | ✓ | × |
| | | 42 | × | × | × | ✓ | × | × | × | × | × |
| | | 43 | ✓ | × | ✓ | ✓ | × | × | × | × | × |
| | | 44 | ✓ | ✓ | × | ✓ | × | × | ✓ | × | × |
| | | 75 | ✓ | ✓ | ✓ | ✓ | × | × | × | ✓ | ✓ |
| | | 76 | ✓ | ✓ | × | × | × | × | × | × | × |
| | | 79 | ✓ | × | ✓ | × | × | × | × | × | × |
| | | 88 | ✓ | × | × | × | × | × | × | × | × |
| | | 90 | ✓ | × | × | ✓ | × | × | × | × | × |
| | | 92 | ✓ | ✓ | × | ✓ | × | × | × | × | ✓ |
| | | 95 | ✓ | ✓ | × | × | × | × | × | ✓ | × |
| | | 96 | ✓ | ✓ | × | ✓ | × | × | × | × | × |
| | | 109 | ✓ | × | × | × | × | × | × | × | × |

| Category | Section | Ref | Relations | | Activities | | | | | Context | |
|---|---|---|---|---|---|---|---|---|---|---|---|
| | | | Fs | Fing | P | RP | FL | M | R | T | CA |
| Diffusion-oriented Approaches | 3.2.2 | 9 | ✓ | ✗ | ✗ | ✓ | ✗ | ✗ | ✗ | ✗ | ✗ |
| | | 10 | ✓ | ✓ | ✓ | ✗ | ✗ | ✗ | ✗ | ✗ | ✗ |
| | | 11 | ✓ | ✗ | ✓ | ✓ | ✗ | ✗ | ✗ | ✗ | ✗ |
| | | 22 | ✓ | ✗ | ✓ | ✓ | ✗ | ✗ | ✗ | ✗ | ✗ |
| | | 33 | ✓ | ✗ | ✗ | ✓ | ✗ | ✗ | ✗ | ✗ | ✓ |
| | | 37 | ✓ | ✗ | ✗ | ✓ | ✗ | ✓ | ✗ | ✗ | ✗ |
| | | 40 | ✗ | ✗ | ✗ | ✓ | ✗ | ✓ | ✓ | ✓ | ✗ |
| | | 41 | ✓ | ✓ | ✓ | ✓ | ✗ | ✗ | ✗ | ✗ | ✗ |
| | | 44 | ✓ | ✓ | ✗ | ✓ | ✗ | ✗ | ✓ | ✗ | ✗ |
| | | 45 | ✗ | ✗ | ✓ | ✗ | ✗ | ✓ | ✗ | ✓ | ✗ |
| | | 77 | ✓ | ✓ | ✓ | ✓ | ✗ | ✗ | ✓ | ✗ | ✓ |
| | | 83 | ✓ | ✗ | ✗ | ✓ | ✗ | ✓ | ✗ | ✗ | ✗ |
| | | 101 | ✓ | ✗ | ✗ | ✓ | ✗ | ✗ | ✗ | ✗ | ✗ |
| | | 104 | ✓ | ✗ | ✗ | ✓ | ✗ | ✗ | ✓ | ✗ | ✗ |
| | | 108 | ✗ | ✗ | ✓ | ✓ | ✗ | ✗ | ✗ | ✗ | ✗ |
| | | 110 | ✗ | ✗ | ✓ | ✗ | ✗ | ✗ | ✗ | ✗ | ✗ |
| | | 126 | ✗ | ✗ | ✓ | ✓ | ✗ | ✗ | ✗ | ✗ | ✗ |

| Category | Section | Ref | Relations | | Activities | | | | | Context | |
|---|---|---|---|---|---|---|---|---|---|---|---|
| | | | Fs | Fing | P | RP | FL | M | R | T | CA |
| Network / Graph Properties | 3.3 | 6 | × | × | × | × | × | ✓ | ✓ | ✓ | × |
| | | 21 | ✓ | × | × | ✓ | × | ✓ | × | × | × |
| | | 50 | ✓ | × | ✓ | × | × | × | × | × | ✓ |
| | | 68 | × | × | ✓ | × | × | × | × | × | ✓ |
| | | 71 | × | × | ✓ | × | × | × | × | × | ✓ |
| | | 76 | ✓ | ✓ | × | × | × | × | × | × | × |
| | | 77 | ✓ | ✓ | ✓ | ✓ | × | × | ✓ | × | ✓ |
| | | 78 | ✓ | ✓ | ✓ | × | × | × | × | × | × |
| | | 80 | ✓ | ✓ | × | × | × | × | × | × | × |
| | | 88 | ✓ | × | × | × | × | × | × | × | × |
| | | 90 | ✓ | × | × | ✓ | × | × | × | × | × |
| | | 100 | ✓ | ✓ | × | × | × | × | × | × | × |
| Ranking | 3.4.1 | 5 | × | × | ✓ | ✓ | × | × | × | ✓ | × |
| | | 19 | ✓ | ✓ | ✓ | ✓ | × | ✓ | × | × | ✓ |
| | | 70 | ✓ | × | ✓ | × | × | × | × | ✓ | × |
| | | 84 | ✓ | × | ✓ | × | × | × | × | ✓ | ✓ |
| | | 86 | ✓ | ✓ | ✓ | ✓ | ✓ | ✓ | × | × | × |
| | | 102 | ✓ | ✓ | ✓ | × | ✓ | × | ✓ | × | × |
| | | 104 | × | × | × | ✓ | × | ✓ | ✓ | × | × |
| | | 111 | ✓ | ✓ | ✓ | ✓ | ✓ | × | × | × | × |
| | | 112 | ✓ | ✓ | ✓ | ✓ | ✓ | × | × | × | × |

| Category | Section | Ref | Relations | | Activities | | | | | Context | |
|---|---|---|---|---|---|---|---|---|---|---|---|
| | | | **Fs** | **Fing** | **P** | **RP** | **FL** | **M** | **R** | **T** | **CA** |
| Recommendation | 3.4.2 | 14 | ✓ | ✓ | ✓ | ✗ | ✗ | ✗ | ✗ | ✗ | ✗ |
| | | 34 | ✓ | ✓ | ✓ | ✗ | ✗ | ✗ | ✗ | ✗ | ✓ |
| | | 35 | ✓ | ✓ | ✓ | ✗ | ✗ | ✗ | ✗ | ✗ | ✓ |
| | | 36 | ✗ | ✓ | ✓ | ✗ | ✗ | ✗ | ✗ | ✓ | ✓ |
| | | 38 | ✓ | ✓ | ✓ | ✗ | ✗ | ✗ | ✗ | ✗ | ✗ |
| | | 63 | ✓ | ✓ | ✗ | ✗ | ✗ | ✗ | ✗ | ✗ | ✓ |
| | | 64 | ✗ | ✗ | ✗ | ✗ | ✗ | ✗ | ✗ | ✗ | ✗ |
| | | 65 | ✓ | ✓ | ✓ | ✗ | ✗ | ✗ | ✗ | ✗ | ✓ |
| | | 66 | ✓ | ✓ | ✓ | ✓ | ✗ | ✓ | ✗ | ✗ | ✓ |
| | | 68 | ✗ | ✗ | ✓ | ✗ | ✗ | ✗ | ✗ | ✗ | ✓ |
| | | 73 | ✓ | ✓ | ✓ | ✓ | ✗ | ✗ | ✗ | ✗ | ✓ |
| | | 75 | ✓ | ✓ | ✓ | ✓ | ✗ | ✗ | ✗ | ✓ | ✓ |
| | | 82 | ✓ | ✓ | ✓ | ✗ | ✗ | ✗ | ✗ | ✗ | ✗ |
| | | 102 | ✓ | ✓ | ✓ | ✗ | ✓ | ✗ | ✓ | ✗ | ✗ |
| | | 103 | ✓ | ✓ | ✗ | ✗ | ✗ | ✗ | ✗ | ✓ | ✗ |
| Other Application Domains | 3.4.3 | 8 | ✓ | ✓ | ✓ | ✓ | ✓ | ✓ | ✗ | ✗ | ✓ |
| | | 28 | ✗ | ✗ | ✓ | ✗ | ✓ | ✗ | ✗ | ✓ | ✗ |
| | | 72 | ✗ | ✗ | ✓ | ✗ | ✗ | ✗ | ✗ | ✗ | ✓ |
| | | 93 | ✓ | ✓ | ✗ | ✗ | ✗ | ✗ | ✗ | ✗ | ✗ |

# 4. Online Social Semantics

In this section, we study the semantics and their role as the second major aspect of OSNs. Specifically, we analyze related works based on Semantic Web technologies along with network theory and graph properties for transforming unstructured data into Linked Data, topic identification, detection of similar users and communities, as well as user personalization (e.g. interests, suggestions, and so on).

## 4.1 Social Modeling

As the adoption of semantics and Linked Data increases, more works have emerged covering aspects of semantic modeling in OSNs. In this section, we present approaches which adopt semantics for modeling the logical topology and structure of social networks and media as well as the information they disseminate.

One of the first studies in this domain is [60], where the use of a specific syntax is proposed for creating a common knowledge representation, by incorporating RDF-like syntaxes into Twitter posts. The use of such statements enables users to freely define relations such as hierarchical or equality relations among hashtags. Hence, an ontology of hashtags is collaboratively created which can be exploited for the resolution of synonymous hashtags or other simple reasoning tasks.

The authors in [29] propose a framework for enriching Twitter messages with semantics relationships by analyzing Twitter posts. These relationships are identified among persons, products, and events and are utilized in order to provide query suggestion to the users.

Another work on the enrichment of Twitter messages with semantics is described in [30]. Specifically, the authors attempt to create user profiles by exploiting Twitter posts by using Semantic Web technologies. In order to capture the users' interests, the URLs of news articles found in tweets are utilized. Lexical analysis is applied on their content so that the relationships between the entities in news articles (representing the interests) can be discovered, which are then semantically related to those tweets.

Social semantics can be exploited in the development of semantic recommender systems. Specifically, the studies [65] and [68], which were analytically presented in Section 3.4.2, propose two semantic followee recommender systems for Twitter. Their aim is to build user interest profiles by exploiting the users' posted messages [65], [68], friendship network [65] and publicly available knowledge bases (i.e. Wikipedia, WordNet, Google) [65], which are then used during the recommendation process.

A framework for inferring user interests in Twitter is also proposed in [69]. In contrast to the ones described above, it is based on the users' followees and the content they consume, rather than their original posts. The proposal is based on the hypothesis that famous people maintain accounts that are being followed by a large number of users. The Wikipedia articles of the former are discovered, linking to a higher level of categories and hierarchies, which become an implicit expression of the users' interests.

The methodology presented in [48] is capable of handling large-scale networks and of generating weighted semantic ones. These networks are created by using comments from a Chinese social network. The methodology focuses on the “giant component” of the derived network in order to reduce the computational complexity so that larger networks can be better handled.

The work described in [55] associates tweets with a given event by utilizing structured information found in them. The initial pool of terms for the retrieval of the messages is manually provided. The final associations take place by applying query expansion techniques and by utilizing the relationships derived by the semantified data.

The authors in [74] create graphs of hashtags found in tweets and utilize their relational information in order to discover latent word semantic connections in cases where words do not co-occur within a specific tweet. Sparseness and noise in tweets are handled by exploiting two types of hashtag relationships: i) explicit ones which refer to hashtags that are contained in a tweet, and ii) potential ones which refer to hashtags that do not appear in a tweet but co-occur with others. Finally, the hashtags and words, which have the highest probability to appear on a specific topic, are discovered.

The following studies employ Semantic Web technologies, ontologies and the DBpedia knowledge base, which is a semantified version of Wikipedia, to achieve their goals.

The study in [31] introduces a semantic data aggregator in Twitter, which combines a collection of compact formats for structured microblog content with Semantic Web vocabularies. Its purpose is to provide user-driven Linked Data. The main focus of this work is on posts and specifically on their creators, content and associated metadata.

Another framework which utilizes semantic technologies, common vocabularies and Linked Data in order to extract microblogging data from scientific events from Twitter, is proposed in [32]. In this work, the authors attempt to identify persons and organization related to them based on geospatial and topic entities.

The authors in [62] propose an ontology-assisted topic modeling technique for determining the topical similarities among Twitter users. The entities found at the posts are mapped to classes of DBpedia ontology, using the DBpedia Spotlight tool, and are used for the labeling of clusters. Moreover, the topical similarities among individuals on different topics are calculated using ranking techniques, which define the structure of the resulting graphs. Based on these graphs, a quasi-clique community detection algorithm is applied for the discovery of topic clusters, without predefining their target number.

Another work using the DBpedia knowledge base is [71], where a framework is proposed for retrieving the context of posts in Twitter by applying graph-based centrality theory. Entities from tweets, in terms of words, are extracted and related to DBpedia URIs in order for semantic concepts to be discovered. Based on the graph centrality theory, a graph of contextualized and weighted entities for each tweet is constructed.

The works in [114] and [117] also attempt to semantically relate Twitter entities to DBpedia URIs. Contrary to the previous work, the entities under investigation are the authors of the posts, and the aim here is to provide a data model of five stars according to Tim Berners-Lee’s Linked Open Data rating system [124]. Furthermore, in [117] a framework is proposed to enable the automatic labeling of Twitter accounts in respect to thematic categories derived from DBpedia properties. Similar to [71], these categories are a result of the semantic concepts of the Twitter-DBpedia interrelation.

In [112] and [114] an ontological schema is presented toward the semantification provision of Twitter social analytics. Specifically, the "InfluenceTracker Ontology" is capable of modeling structural aspects of OSNs including the accounts, the users owning them, their disseminated entities (mentions, replies, hashtags, photos, and URLs), and friendship relationships. Moreover, the representation of qualitative metrics such as the influence of the accounts or various likability and impact metrics is also possible.

## 4.2 Social Matching

The studies presented in this section exploit the use of social semantics for identifying similar properties and activities with respect to user-generated content, description of real-life events, as well as revealing user interests and behavioral patterns across different online social media users. Thus, we divide this topic into two subtopics, namely a) User-oriented (e.g. similar user recommendation, user preferences etc.), and b) Topic and Event-oriented (e.g. topic profiling and user interest, event detection, product marketing and others).

### 4.2.1 User-oriented Matching

Despite the fact that the set of social semantics of each account in an OSN is unique as they depend on personal social activities, common patterns among them can be recognized. These can be exploited to enable the discovery of users' social behavior and preferences.

The study presented in [1] describes a framework using supervised learning for distinguishing users in OSNs according to their influence and revealed the communities they belong to. The authors do not define a new influence measure, but discovered predictive properties associated with the users' activity level and involvement in those communities. The supervised learning is based on follow-up relationships, interactions (mentions, replies), the structure and activity of the network, the centrality of users, and the quality of the tweets. The study concludes that those relationships are the most important for identifying influential users.

The aim in [4] is also the identification of influential users based on their interactions in their posts on a given topic. Toward this end, a graph model representing the relationships of the posts is created, which is then transformed into a user graph. The latter is used for the discovery of influential users by considering properties and measures from both graphs. Similarly, as described in Section 3.1.2, the authors in [123] follow the same approach and apply the PageRank algorithm on that graph for the detection of influencers. The posts belonging to a specific topic are discovered through a fuzzy word similarity algorithm which utilized all the contents of the messages.

The work described in [5] was presented in Section 3.4.1. Influencers are regarded as those generating tweets of high quality. Their quality is evaluated according to a set of parameters such as the topic focus degree, the retweeting behavior, and the topic-specific influence of the users who retweeted those messages.

The framework proposed in [6], as presented in Section 3.3, aims identify influential users in topic-based communities. A measure of alpha centrality is employed on a graph derived from direct communications, which incorporates both directionality of network connections and a measure of external importance.

Influential users are discovered in [13] by applying an algorithm as an extension of PageRank, which takes into consideration both the topical similarity among users and their link structure. It is claimed that due to homophily, that is the tendency of individuals to associate and bond with others having similar interests, most of the "Follower-Following" relations appear. This work also suggests that the active users are not necessarily influential.

In Section 3.4.1 we described a study ([84]) where the problem of topic-sensitive opinion leaders' identification in online review communities is investigated. Toward this end a two-staged approach is presented. Initially the opinion leaders' expertise and interests are derived from their tags found at the description of the products. During the next stage a computational approach measures the leaders' influence and ranks them according to not only the link structure of customer networks but also to their expertise and interests. The influence depends on the topical similarity among reviewers on a specific topic.

The task of the topic experts' identification, namely influential users on specific domains, is also presented in [16]. A post-feature based approach is proposed which utilizes nine kinds of features reflecting how the users interact. Their aggregation results in the production of three different kinds of influence measurements.

The authors in [20] claim that a user's influential level can be detected by considering the writing style and behavior within the OSNs. Therefore, they proposed 23 features of user profile (e.g. presence of hashtags, URLs, self-mentions, number of followers and tweets) and 9 features of tweets (e.g. extension, frequency, quality, number of retweets) that can affect influence impact. By applying machine learning algorithms, the most influential users are identified.

A framework-exploiting machine learning techniques for discovering top persuasive users in OSNs is described in [99]. The proposed persuasiveness metric is pair-wise and is based on three factors: influence, entity similarity, and structural equivalence. Influence depends on the strength of social interactions between users. Entity similarity measures how close two profiles are. Structural equivalence measures the structural similarity of two entities according to a distance function. Each of these factors is assigned a probability which denotes the likelihood of persuasion. A machine-learning algorithm achieves the prediction of these probabilities.

The rebroadcasting behavior of users in OSNs is studied in [94]. Specifically, a model is proposed which examines three aspects: the role of content, content-user fit, and social influence. The "content-user fit" measure considers the interaction between the message content and user interests. As in [91], influence measures the susceptibility of users for identifying those whose posts affect the reposting behavior of others. In order to discover the users' interests, the well-known Latent Dirichlet Allocation (LDA) [59] methodology is applied on each message. The study concludes that the rebroadcasting of messages does not depend only on its content but also on its relevance with a user.

The LDA topic modeling approach is also applied in [67], where a user centric topic discovery framework is proposed. The users' tweets are analyzed for identifying their interests and for creating personalized topic profiles. Toward this end a Part-Of-Speech (POS) tagger extracts the nouns of the tweets, which are provided to a search engine to retrieve the top documents based on their relevance. Using LDA on the content on those web pages the final topics are provided.

Another framework, which was also described in another section (Section 4.1) of this survey, is employed for inferring user interests in Twitter [69]. Contrary to the previous frameworks, it is based on the users' followees and the content they consume, rather than their original posts. The proposal is based on the hypothesis that famous people maintain accounts being followed by a large number of users. The Wikipedia articles of the former are discovered, linking to a higher level of categories and hierarchies, which become an implicit expression of the users' interests.

The following studies exploit the social semantics in OSNs in order to propose query expansion techniques for providing an enriched coverage of information needs.

The study in [56] describes a query expansion framework that takes into account the users' preferences which are derived by analyzing microblog posts and hashtags related to the targeted users.

Another query expansion approach is proposed in [57]. It takes into consideration the similarity between tags composing a query and the social proximity between the query and the user's profile. Its aim is to assist users by refining and formulating their queries and by providing them with information relevant to their interests.

The research effort of [55] we have presented in Section 4.1, attempts to associate tweets with a given event, by utilizing their structured information. The application of query expansion techniques and the relationships derived from the semantified data result in those associations.

The story-tracking framework of study [72] is modeled as a pattern mining and real-time retrieval problem. The most popular news stories, assigned with hashtags, are detected by mining frequent hashtag pattern sets. Using query expansion on the original hashtags new story articles are retrieved. The pattern set structure enables hierarchical and multiple-linkage representation of the articles.

The authors in [58] attempt to identify several hashtags relevant to a given query, that can be used to expand it thus leading to more accurate content retrieval. The proposed method leverages statistical techniques to build probabilistic language models for each available hashtag through a suitable microblog posts corpus.

Another study in the area of query expansion with respect to a user's input query is introduced in [115] and [116]. Specifically, the authors proposed an algorithmic approach that can create a dynamic query suggestion set, which consists of the most viral and trendy Twitter entities (hashtags, user mentions and URLs) when considering the initial input query. In contrast to the work in [58], apart from hashtags, the expansion set may also contain other accounts and URLs resulting in a more enriched coverage of information needs. Moreover, the framework is not based solely on statistical techniques and probabilistic models but on the capture-recapture methodology [127], thus making the set dynamic by providing a survival probability on the entities. The capture-recapture experiments are used in wildlife biological studies, where subjects of investigation are captured, marked and then released. If a marked individual is captured on a subsequent trapping occasion, then it is mentioned as "recaptured". Based on the number of recaptured individuals, we can estimate the total population size, as well as the birth, death and survival rate of each species under study.

In Section 4.1 several studies ([29], [30], [32], and [62]) utilize semantic technologies and related protocols to provide expanded query suggestions or to represent user preferences and similarities. Specifically, the authors of [29] proposed a framework for enriching Twitter messages with semantic relationships by analyzing Twitter posts. These relationships are identified among persons, products, and events and are utilized in order to provide query suggestion to the users. The authors attempt to identify persons and organization related to them based on geospatial and topic entities. The study in [59] uses Semantic Web technologies for the creation of user profiles by analyzing Twitter posts. In order to capture the users' interests, the URLs of news articles found in tweets are used. Lexical analysis is applied on their content in order to discover the relationships between the entities in news articles (representing the interests) which are then semantically related to those tweets. The framework in [32] exploits common vocabularies and Linked Data in order to extract microblogging data regarding scientific events from Twitter. Finally, an ontology-assisted topic modeling technique for determining the topical similarities among Twitter users is proposed in [62]. The entities found at the posts are mapped to classes of DBpedia ontology and are used for the labeling of clusters. Moreover, the topical similarities among individuals on different topics are calculated using ranking techniques, which define the structure of the resulting graphs. Based on these graphs, a quasi-clique community detection algorithm is applied for the discovery of topic clusters, without predefining their target number.

In Section 3.3 and 3.4.2, we presented studies that attempt to adequately describe user characteristics, in order for similarities among them to be discovered. In [38] a matrix factorization framework with social regularization is proposed for improving recommender systems by incorporating social network information. Each social link is weighted based on the similarity among the users, allowing the exploitation of friends differently according to the rating similarity.

The friend recommendation problem in Flickr is studied in [63], from the viewpoint of network correlation. The authors assume that each user has many different social roles in OSNs. During each role different social sub-networks are formed which are aligned in order for the correlations among them to be found through a weighted tag feature selection. When recommendations are made, the similarities of the tag features among the new and existing users are calculated. The more similar the tags are, the closer the users should be.

The author in [68] proposes a semantic follower recommender system in Twitter which exploits the users' tweets in order to build interest profiles. An interest graph is created using specific semantic knowledge graphs containing a variety of topics, which are then mapped to the users according to their semantic relevance to the topics. Using graph theory algorithms user interest similarity is calculated which is used during the recommendation process.

A framework for discovering similar accounts in Twitter based only on the "List" feature is proposed in [64]. This functionality allows the users to create their own lists by adding any account they wish. The authors claim that this feature is considered a form of crowdsourcing. The hypothesis of the methodology is that when two accounts are present in the same list they should be similar or related to each other. Therefore, the proposed measurement relies on the number of lists that a specified account and a potentially similar one are listed together.

Another study on the discovery and suggestion of similar Twitter users is described in [113]. It is based entirely on their disseminated content in terms of Twitter entities used (mentions, replies, hashtags, URLs). The framework is based on the hypothesis that many of these entities are used frequently by multiple accounts. Consequently, the more common entities are found in the messages of different accounts, the more similar, in terms of content or interest, they tend to be. The methodology is based exclusively on semantic representation protocols and related technologies.

### 4.2.2 Topic and Event-oriented Matching

As we have already mentioned, social semantics patterns can be used to identify users' interests or topics of discussion such as real-life events.

The studies described in [23], [4], [5], and [95] are specialized in discovering the most influential authors in Twitter on a specific topic. Specifically, in [23], the authors suggest a set of metrics based on original tweets, replies, retweets, mentions and friendship relationships. In [4], which was also described in Section 4.2.1, these metrics are discovered by considering properties and measures on user-post graphs, while in [5], presented in Section 3.4.1, influencers are regarded as those generating tweets of high quality. A different kind of social influence, a persuasive one, is proposed in [95]. The proposed measurement depends on topical information, the users' authority and the characteristics of relationships among individuals. Section 3.2.1 presents more detailed information on the study.

The authors in [92] propose a content-centered model of flow analysis for investigating the Influence Maximization problem on topic-specific influencers. As also described in Section 3.2.1, this analysis is not based on the users' relationships, but on the content of the transmitted messages. Influencers are discovered by exploiting information flow patterns, their position, as well as the number of flow paths they participate in.

A framework for determining the relevance of Twitter messages for a given topic is introduced in [54]. Two feature categories are identified, i.e., features related to the user query and thus calculated as soon as the latter is formed, and features that are not related to this query but are inherent posts and are therefore calculated when they are modified.

In Section 4.1, we presented an approach that associates tweets with a given event, by utilizing their structured information [55]. The application of query expansion techniques and the relationships are derived from the semantified data result in those associations.

Another topic-oriented framework for Twitter is presented in [51]. Its aim is to discover the users' topics of interest by examining the entities found in their posts, which may be mentions or plain text (in OSNs the mentions are words prefixed with "@"). The Wikipedia knowledge base is leveraged in order to disambiguate those entities and the topics of interest to be defined (e.g. the term “apple” may refer to the fruit or to the multinational technology company).

The work in [67] was described in Section 4.2.1 and presents an LDA [59] topic profile modeling approach for the discovery of users’ interests. Therefore, a POS tagger extracts the nouns from their tweets, which are then provided to a search engine to retrieve the top relating web pages. Using LDA on the content of these web pages discovers final topics.

Topic profiling using the Wikipedia knowledge base is also studied in [52]. Specifically, the topics are discovered based on the posted content, namely hashtags, of Twitter accounts and their friendship relationships. The celebrities (accounts of popular people) who are followed are the primary indicators of interest which have been derived from their Wikipedia classification. The indicators along with the posted hashtags infer the topics of interest of the accounts.

Similarly, the work in [53] focused on extracting the interests of Twitter accounts based on their generated messages. The methodology leverages the hierarchical relationships found in Wikipedia in order to infer user interests. The authors claim that the hierarchical structures can improve existing systems to become more personalized based on broader and higher level concepts (e.g. the concept “Basketball” is more generic that the term “NBA”).

The authors in [62] proposed an ontology-assisted topic modeling technique for determining the topical similarities among Twitter users. The entities found in the posts are mapped to classes of DBpedia ontology and are used to label clusters. Moreover, the topical similarities between individuals on different topics are calculated using ranking techniques, which define the structure of the resulting graphs. Based on the graphs, a quasi-clique community detection algorithm is applied for discovering topic clusters without predefining their target number. The methodology was also presented in Section 4.1.

Another topic-oriented framework, also presented in Section 3.3, which uses the DBpedia knowledge base is proposed in [71]. Specifically, the context of Twitter is mapped to DBpedia entities and graph-based centrality theory is applied for assigning weights to the entities of the examined messages.

Topic profiling is also exploited in recommendation systems and examples of such studies have been proposed in [65], [68], and [103], which were also presented in Section 3.4.2. The first two ([65] and [68]) describe methodologies for the creation of semantic followee recommender systems for Twitter. These studies are based on the classification of the content of tweets and the users that generated them and on semantic knowledge graphs containing a variety of topics being mapped to users respectively. The recommender described in [103] discovers personal interests by applying a distributed learning supervised algorithm and by taking into consideration explicit social features such as the users’ topic-level influence, topic information, and social relations.

A framework for discovering topic-specific experts in Twitter by employing two distinct metrics is presented in [70]. The first metric measures the users’ global authority on a given topic, while the other metric provides the similarity between the users’ generated tweets and that topic. By leveraging the topical influence and similarity, the users who have the highest-ranking scores are regarded as experts in that domain. This study is described in detail in Section 3.4.1.

In Section 3.2.1, we described a multi-topic influence propagation model based on user relationships, posts and social actions [75]. The influence score consists of direct and indirect influence, related to different topics. The distribution of users' topics of interest depends on the content of the disseminated messages. The proposed topic-dependent algorithm is applied and a multi-topical network is created in order to identify multi-topic influential users.

Another framework exploiting both user interests and social influence is presented in [94]. Specifically, the rebroadcasting behavior of users in OSNs is studied. The proposed model examines three aspects; the role of content, content-user fit, and social influence. The "content-user fit" measure considers the interaction between the message content and user interests. This study was also presented in Section 4.2.1. In the same Section, study [1] presented a framework using supervised learning for discovering the communities the users belong to and identifying the most influential ones.

## 4.3 Community Detection

Community detection is not only useful for the analysis of OSNs but also for understanding the structure and properties of complex networks. The aim is to group their nodes into potentially overlapping sets, sharing common attributes and characteristics. The following studies propose different approaches to detect communities in OSNs.

In contrast to the traditional techniques, the approach in [47] uses both the structure of the network and the attributes of the nodes in order to develop an algorithm for detecting overlapping communities.

Another algorithm for detecting communities is presented in [49]. It is based on the content of the edges derived from the users' pair wise interactions. According to the authors, this algorithm provides richer insights into communities because it depicts the nature of the interactions more effectively.

As described in Section 4.1, the methodology in [48] aims at discovering the latent communities in large-scale networks and generates weighted semantic ones. The latter are created using the information extracted from users' comment content.

The approach in [50] was presented in Section 4.1 and detects communities using node similarity techniques. Specifically, a virtual network is created, where virtual edges are inserted into the original network based on the similarity of the nodes which is calculated using the Jaccard Measure. The proposed algorithm is then applied on the generated virtual network.

In the same section, the study [62] proposes a topic modeling technique among Twitter users using the DBpedia ontology. A community detection algorithm is applied on the users' graph to discover the topics.

Community detection can also be used for the identification of influential users in OSNs. Specifically, the work in [76] proposes such a methodology by applying maximum flow algorithms on a weighted representation of the network by considering structural features such as shortest path, betweeness, closeness and degree centralities. More information can be found in Section 3.2.1.

Another framework, also presented in Section 4.2.1, for discovering top persuasive users in OSNs is described in [99]. The framework is based on machine learning techniques and depends on three factors: influence, entity similarity, and structural equivalence. Each of them is assigned a probability (denoting the likelihood of persuasion) that has been derived by using a machine-learning algorithm.

Another study that uses communities to identify influential users is presented in [27]. The authors analyzed the social activity and the interconnections of the users inside the communities they belong to. The communities are utilized to develop a framework for modeling the spread of influence by identifying the most influential users.

## 4.4 Comparison of Related Works

Similarly to Section 3.5, we provide here comparative insights (Table 4) from the above reviewed articles that refer to online social semantics. Moreover, for each reviewed article, the first three columns denote its category according to our classification scheme (see Figure 5), the section number, as well as its reference number. In the rest columns, we place a mark of "Yes" (✓) or "No" (✗), in case where the studies employ or propose metrics and characteristics based on:

a. Network Structure (NS): includes social follow-up relationships or other types of network linking (e.g. based on mention, reply actions).
b. Behavioral/Conversational activities: posts (P), re-posts (RP), favorites/likes (FL), mentions (M), replies (R), Contextual analysis: works for building user profiles or providing personalized information (P), and those applied on specific topics (T).
c. Use of Knowledge Bases (KB): works that use publicly available, open or crowdsourcing-based resources (e.g. Wikipedia, DBpedia, and so on).
d. Use of semantics: modeling unstructured data using RDF protocol, use of (existing or new) ontologies (O).
e. OSN Entities: hashtags (H) and web URLs distributed in social content.

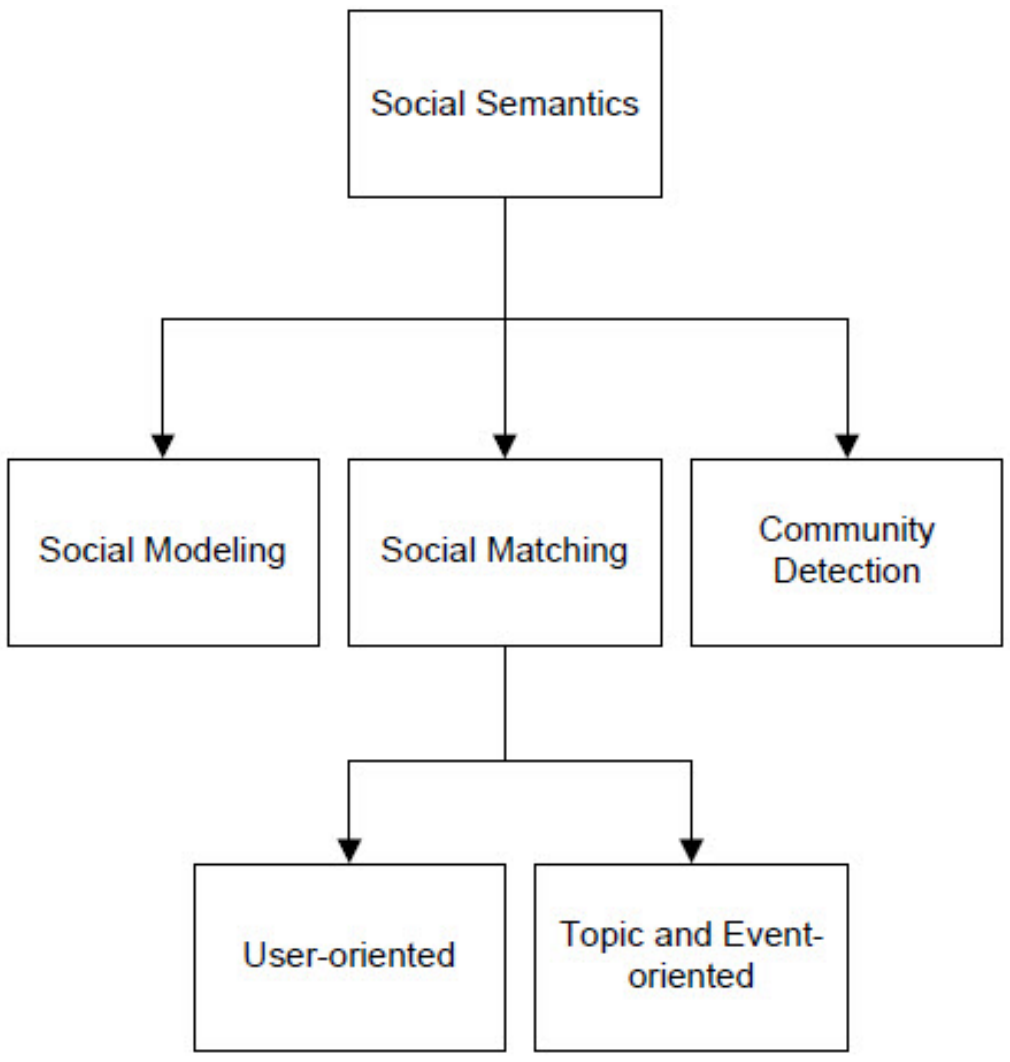

Figure 5: The hierarchical classification scheme followed in this work on Social Semantics [part of Figure 1]

Table 4: Classification of referenced works. NS refers to network structure, (R)P to (re)-posts, I to interactions, P to profiling and personalization, T to topics, O to ontologies, and H to hashtags.

| Category | Section | Ref | NS | Activities | | Context | | KB | Semantic Web | | Entities | |
|---|---|---|---|---|---|---|---|---|---|---|---|---|
| | | | | (R)P | I | P | T | | RDF | O | H | URLs |
| Social Modeling | 4.1 | 29 | ✗ | ✓ | ✗ | ✗ | ✗ | ✓ | ✗ | ✗ | ✗ | ✗ |
| | | 30 | ✗ | ✓ | ✗ | ✓ | ✗ | ✗ | ✓ | ✗ | ✓ | ✓ |
| | | 31 | ✗ | ✓ | ✗ | ✗ | ✗ | ✗ | ✓ | ✓ | ✓ | ✓ |
| | | 32 | ✗ | ✓ | ✗ | ✗ | ✓ | ✓ | ✓ | ✓ | ✗ | ✗ |
| | | 48 | ✓ | ✓ | ✓ | ✗ | ✗ | ✗ | ✗ | ✗ | ✗ | ✗ |
| | | 55 | ✗ | ✓ | ✗ | ✗ | ✓ | ✓ | ✓ | ✗ | ✗ | ✗ |
| | | 60 | ✗ | ✓ | ✗ | ✗ | ✗ | ✗ | ✓ | ✗ | ✓ | ✗ |
| | | 62 | ✓ | ✓ | ✗ | ✗ | ✓ | ✓ | ✗ | ✓ | ✗ | ✗ |
| | | 65 | ✓ | ✓ | ✗ | ✓ | ✗ | ✓ | ✗ | ✗ | ✗ | ✗ |
| | | 68 | ✗ | ✓ | ✗ | ✓ | ✗ | ✓ | ✗ | ✗ | ✗ | ✗ |
| | | 69 | ✓ | ✓ | ✗ | ✓ | ✗ | ✓ | ✗ | ✗ | ✗ | ✗ |
| | | 71 | ✓ | ✓ | ✗ | ✗ | ✓ | ✓ | ✗ | ✗ | ✗ | ✗ |
| | | 74 | ✗ | ✓ | ✗ | ✗ | ✓ | ✗ | ✗ | ✗ | ✓ | ✗ |
| | | 112 | ✓ | ✓ | ✓ | ✗ | ✗ | ✗ | ✓ | ✓ | ✓ | ✓ |
| | | 114 | ✓ | ✓ | ✓ | ✗ | ✗ | ✓ | ✓ | ✓ | ✓ | ✓ |
| | | 117 | ✗ | ✗ | ✗ | ✓ | ✗ | ✓ | ✗ | ✗ | ✗ | ✗ |

| Category | Section | Ref | NS | Activities | | Context | | KB | Semantic Web | | Entities | |
|---|---|---|---|---|---|---|---|---|---|---|---|---|
| | | | | (R)P | I | P | T | | RDF | O | H | URLs |
| User-oriented | 4.2.1 | 1 | ✓ | ✓ | ✓ | ✗ | ✗ | ✗ | ✗ | ✗ | ✗ | ✗ |
| | | 4 | ✓ | ✓ | ✗ | ✗ | ✓ | ✗ | ✗ | ✗ | ✗ | ✗ |
| | | 5 | ✓ | ✓ | ✗ | ✗ | ✓ | ✗ | ✗ | ✗ | ✗ | ✗ |
| | | 6 | ✓ | ✗ | ✓ | ✗ | ✓ | ✗ | ✗ | ✗ | ✗ | ✗ |
| | | 13 | ✓ | ✗ | ✗ | ✗ | ✓ | ✗ | ✗ | ✗ | ✗ | ✗ |
| | | 16 | ✓ | ✓ | ✗ | ✗ | ✓ | ✗ | ✗ | ✗ | ✗ | ✗ |
| | | 20 | ✗ | ✓ | ✓ | ✗ | ✗ | ✗ | ✗ | ✗ | ✓ | ✓ |
| | | 29 | ✗ | ✓ | ✗ | ✗ | ✗ | ✓ | ✗ | ✗ | ✗ | ✗ |
| | | 30 | ✗ | ✓ | ✗ | ✓ | ✗ | ✗ | ✓ | ✗ | ✓ | ✓ |
| | | 32 | ✗ | ✓ | ✗ | ✗ | ✓ | ✓ | ✓ | ✓ | ✗ | ✗ |
| | | 38 | ✓ | ✓ | ✗ | ✗ | ✗ | ✗ | ✗ | ✗ | ✗ | ✗ |
| | | 55 | ✗ | ✓ | ✗ | ✗ | ✓ | ✓ | ✓ | ✗ | ✗ | ✗ |
| | | 56 | ✗ | ✓ | ✗ | ✓ | ✗ | ✗ | ✗ | ✗ | ✓ | ✗ |
| | | 57 | ✗ | ✗ | ✗ | ✓ | ✗ | ✗ | ✗ | ✗ | ✗ | ✗ |
| | | 58 | ✗ | ✓ | ✗ | ✗ | ✗ | ✗ | ✗ | ✗ | ✓ | ✗ |
| | | 62 | ✓ | ✓ | ✗ | ✗ | ✓ | ✓ | ✗ | ✓ | ✗ | ✗ |
| | | 63 | ✓ | ✗ | ✗ | ✗ | ✓ | ✗ | ✗ | ✗ | ✗ | ✗ |
| | | 64 | ✓ | ✗ | ✗ | ✗ | ✗ | ✗ | ✗ | ✗ | ✗ | ✗ |
| | | 67 | ✗ | ✓ | ✗ | ✓ | ✗ | ✗ | ✗ | ✗ | ✗ | ✓ |
| | | 68 | ✗ | ✓ | ✗ | ✓ | ✗ | ✓ | ✗ | ✗ | ✗ | ✗ |
| | | 69 | ✓ | ✓ | ✗ | ✓ | ✗ | ✓ | ✗ | ✗ | ✗ | ✗ |
| | | 72 | ✗ | ✓ | ✗ | ✗ | ✗ | ✗ | ✗ | ✗ | ✓ | ✗ |
| | | 84 | ✓ | ✓ | ✗ | ✓ | ✗ | ✗ | ✗ | ✗ | ✗ | ✗ |

| Category | Section | Ref | NS | Activities | | Context | | KB | Semantic Web | | Entities | |
|---|---|---|---|---|---|---|---|---|---|---|---|---|
| | | | | (R)P | I | P | T | | RDF | O | H | URLs |
| User-oriented | 4.2.1 | 91 | ✗ | ✓ | ✓ | ✗ | ✗ | ✗ | ✗ | ✗ | ✗ | ✗ |
| | | 94 | ✗ | ✓ | ✗ | ✓ | ✗ | ✗ | ✗ | ✗ | ✗ | ✗ |
| | | 99 | ✓ | ✗ | ✓ | ✗ | ✗ | ✗ | ✗ | ✗ | ✗ | ✗ |
| | | 113 | ✗ | ✓ | ✓ | ✗ | ✗ | ✗ | ✓ | ✓ | ✓ | ✓ |
| | | 115 | ✗ | ✓ | ✓ | ✗ | ✗ | ✗ | ✗ | ✗ | ✓ | ✓ |
| | | 116 | ✗ | ✓ | ✓ | ✗ | ✗ | ✗ | ✗ | ✗ | ✓ | ✓ |
| | | 123 | ✓ | ✓ | ✓ | ✗ | ✓ | ✗ | ✗ | ✗ | ✓ | ✗ |
| Topic and Event-oriented | 4.2.2 | 1 | ✓ | ✓ | ✓ | ✗ | ✗ | ✗ | ✗ | ✗ | ✗ | ✗ |
| | | 4 | ✓ | ✓ | ✗ | ✗ | ✓ | ✗ | ✗ | ✗ | ✗ | ✗ |
| | | 5 | ✓ | ✓ | ✗ | ✗ | ✓ | ✗ | ✗ | ✗ | ✗ | ✗ |
| | | 23 | ✓ | ✓ | ✓ | ✗ | ✓ | ✗ | ✗ | ✗ | ✓ | ✓ |
| | | 51 | ✗ | ✓ | ✓ | ✓ | ✓ | ✓ | ✗ | ✗ | ✗ | ✗ |
| | | 52 | ✓ | ✓ | ✗ | ✓ | ✗ | ✓ | ✗ | ✗ | ✓ | ✗ |
| | | 53 | ✗ | ✓ | ✗ | ✓ | ✗ | ✓ | ✗ | ✗ | ✗ | ✗ |
| | | 54 | ✓ | ✓ | ✓ | ✗ | ✓ | ✓ | ✗ | ✗ | ✓ | ✓ |
| | | 55 | ✗ | ✓ | ✗ | ✗ | ✓ | ✓ | ✓ | ✗ | ✗ | ✗ |
| | | 62 | ✓ | ✓ | ✗ | ✗ | ✓ | ✓ | ✗ | ✓ | ✗ | ✗ |
| | | 65 | ✓ | ✓ | ✗ | ✓ | ✗ | ✓ | ✗ | ✗ | ✗ | ✗ |
| | | 67 | ✗ | ✓ | ✗ | ✓ | ✗ | ✗ | ✗ | ✗ | ✗ | ✓ |
| | | 68 | ✗ | ✓ | ✗ | ✓ | ✗ | ✓ | ✗ | ✗ | ✗ | ✗ |
| | | 70 | ✓ | ✓ | ✗ | ✗ | ✓ | ✗ | ✗ | ✗ | ✗ | ✗ |
| | | 71 | ✓ | ✓ | ✗ | ✗ | ✓ | ✓ | ✗ | ✗ | ✗ | ✗ |
| | | 75 | ✓ | ✓ | ✓ | ✗ | ✓ | ✗ | ✗ | ✗ | ✗ | ✗ |

| Category | Section | Ref | NS | Activities | | Context | | KB | Semantic Web | | Entities | |
|---|---|---|---|---|---|---|---|---|---|---|---|---|
| | | | | (R)P | I | P | T | | RDF | O | H | URLs |
| Topic and Event-oriented | 4.2.2 | 92 | ✓ | ✓ | ✗ | ✗ | ✗ | ✗ | ✗ | ✗ | ✗ | ✗ |
| | | 94 | ✗ | ✓ | ✗ | ✓ | ✗ | ✗ | ✗ | ✗ | ✗ | ✗ |
| | | 95 | ✓ | ✗ | ✗ | ✗ | ✓ | ✗ | ✗ | ✗ | ✗ | ✗ |
| | | 103 | ✓ | ✓ | ✗ | ✓ | ✗ | ✗ | ✗ | ✗ | ✗ | ✗ |
| Community Detection | 4.3 | 27 | ✓ | ✓ | ✗ | ✗ | ✗ | ✗ | ✗ | ✗ | ✗ | ✗ |
| | | 47 | ✓ | ✗ | ✗ | ✗ | ✗ | ✗ | ✗ | ✗ | ✗ | ✗ |
| | | 48 | ✓ | ✓ | ✓ | ✗ | ✗ | ✗ | ✗ | ✗ | ✗ | ✗ |
| | | 49 | ✓ | ✗ | ✗ | ✗ | ✗ | ✗ | ✗ | ✗ | ✗ | ✗ |
| | | 50 | ✓ | ✗ | ✗ | ✗ | ✗ | ✗ | ✗ | ✗ | ✗ | ✗ |
| | | 62 | ✓ | ✓ | ✗ | ✗ | ✓ | ✓ | ✗ | ✓ | ✗ | ✗ |
| | | 76 | ✓ | ✗ | ✗ | ✗ | ✗ | ✗ | ✗ | ✗ | ✗ | ✗ |
| | | 99 | ✓ | ✗ | ✓ | ✗ | ✗ | ✗ | ✗ | ✗ | ✗ | ✗ |

# 5. Modeling quality content in OSNs

In sections 3 and 4, we studied the impact of influence and the role of semantics in OSN analysis. Their combination can be used for assessing information dynamics as well as for the qualitative assessment of viral user-generated content. Here, we highlight the key points and considerations towards the definition and semantification of quality user-generated content.

According to the authors in [24], metrics regarding retweets are the best quantitative indicators that show a preference for reading a tweet over another. From the readers' perspective, a tweet being retweeted several times is more attractive than a tweet with a lot of mentions. The authors conclude that the relationship among users and authors is the best qualitative indicator which has the strongest effect on the retweeting and reading processes. Retweets as quality indicators are also considered for measuring the appreciation of other users on the generated posts [1]. This attribute is highly used to calculate users' influence.

The study in [26] suggests that retweeting can also be characterized as a conversational frame. According to the authors, a conversation "exists" either during a retweet where some new information can be added to the initial message, or when a single tweet is retweeted multiple times. The latter is interpreted as an action to invite new users into the conversation.

The rebroadcasting behavior of users is also studied in [94] by examining three aspects: the role of content, content-user fit, and social influence. Content-user fit considers the interaction between the message content and user interests whereas social influence measures the impact on users' re-posting behaviors.

The work in [66] describes a service for proposing news articles to Twitter users. It is based not only on terms, which are mined from the users' and their friends' timeline, but it also incorporates additional factors that affect the interest of a user on a tweet, the number of retweets, and the influence of its publisher.

Personalization issues for recommendation purposes are also examined in [103]. The authors incorporate influential features (e.g. users' topic-level influence, topic information) and their relations among OSN users (e.g. retweeting behavior) for improving recommendation results in thematic categories. In addition, social influence and its propagation can also be used as quality indicators in recommendation systems. In the work described in [14], influence is considered as an attribute that propagates among users in OSNs. The authors' proposed framework calculates the influence that social relationships have on users' rating behaviors, and incorporates it into recommendation proposals.

A study that evaluates the quality of articles on Wikipedia by investigating their usage on Twitter is presented in [61]. This is achieved by analyzing three aspects, namely the language used in tweets in their referenced Wikipedia articles, the Twitted-related content features of such articles (e.g. URLs, hashtags, names of entities), and the correlation between the number of tweets/retweets and edits in their related articles. The authors discovered that the language of the tweets and the referenced Wikipedia articles are not always the same, mainly because of the low quality or the absence of equivalent entries in the user's native language. Moreover, it was found that the impact of a tweet/retweet about a certain topic is not related to crowdsourcing-based metrics (e.g. edits, discussions, etc.) on the same Wikipedia topic.

In Section 2, we also described a framework [118] that exploits influence for evaluating and enhancing communication issues between governmental agencies and citizens (OSNs users). The authors evaluate the quality of the agencies' responses with respect to the citizens' requests, analyze their sentiment and suggest influential users for agencies in order to obtain a new audience.

The authors in [106] analyze and compare a variety of measurements in OSNs that affect user influence. These were grouped under various criteria, namely neighborhood (i.e. number of influencers, personal network exposure), structural diversity (i.e. active community metrics), locality, temporal measures (i.e. retweet time delay), cascade measures (i.e. size, path length), and metadata (i.e. presence of links, mentions, hashtags). Moreover, based on several learning algorithms the authors propose methods to calculate the users' retweeting probability.

Another interesting methodology for measuring user influence based on the content quality is proposed in [89]. Initially, users who disseminate quality content are considered those with high Follower-to-Followee ratio. According to the classification methodology, in the case of spam detection the users' influence is reduced. The authors introduced time as an important factor that affects the content influence and its probability of being viewed, retweeted or commented in different time zones.

Influential users are discovered in [13] by applying an algorithm as an extension of PageRank, which takes into consideration both the topical similarity between users and their link structure. It is claimed that due to homophily (i.e. the tendency of individuals to associate and bond with others having similar interests) most of the "Follower-Following" relations appear.

Another approach that defines influence according to the behavior of our directly related users (e.g. friends, followers, and so on) is presented in [25]. The authors proposed an "influenceability" score aimed at representing a user's susceptibility to be influenced by others. It is built on the hypothesis that very active users perform actions without getting influenced by anyone. The study concluded that such kind of users should be considered as influencers in a network. Following the same rationale, in [39], the authors investigated social influence as a way of dictating users' behaviors in order to impose similar behavior.

The authors in [81] identified influential users by using association rule learning. As a machine-learning technique the specific approach investigates how one item affects another by analyzing the frequency and simultaneous appearance of certain items in a specific dataset. The authors assessed user participation in a post based on previous interactions with other users on common posts. Toward this end, posts from Facebook pages were analyzed by extracting users' actions, such as comments and likes. The authors claim that this technique allows the prediction of the participation of a particular user on a post discussion based on other users' activities.

In [93], the authors investigated whether peer and external influence can be inferred by using the user's friendship network. The experiment took place during an on-line voting procedure in Facebook. The analysis of the users' demographics and votes showed strong homophily among the communities and friends' votes. The activation of voters propagated from recently activated friends and news agencies.

In [95] the authors analyzed the so-called persuasion-driven social influence based on topic. Specifically, several influence measurements in terms of influence propagation for quantifying user-to-user influence probability incorporate users' social persuasiveness. Based on the proposed metrics, the framework exploits the topical information, the user's authority and the characteristics of relationships (such as direct or indirect connections among users) among individuals.

In [105], the authors proposed an influence assessment approach for OSNs, by addressing limitations such as the lack of combined relationships and uncertainty ignorance of existing ones. An influence graph is created to enable the observation of different relations and interactions, including retweets, mentions, and replies. Based on the belief functions theory, a general influence measure for a given user is established by information fusion of the different relations. The proposed influence measure takes into account different interaction patterns in the graph, and considers derives influence from indirect nodes.

In Section 2, we presented the "Influence Metric" measurement [111] and its improved version [112] in the calculation of the importance and influence of Twitter accounts. Therefore, a social function is presented incorporating the activity (or passivity) of a user, and the number of followers and followees. In the improved version a new qualitative factor is incorporated into the measurement based on the established *h*-index measurement. Its aim is to reflect other users' actions and preferences over the content of the created posts, thus enhancing the influence of quality users.

Finally, an important aspect in modeling the content quality in OSNs is the credibility of users because it is also strongly related to the trustworthiness of the information itself. One proposed solution is crowd-based neutralized evaluation based on the accuracy, clarity and timeliness of information since its creation. When user-generated content is created, topic/domain labeling enhances its credibility. In addition to personalization provision, each OSN user should also be able to create personalized filtering rules, in order to select and evaluate the labeled content. By employing this mechanism, the community will eventually evaluate the topic/domain label credibility through collective intelligence and crowd-sourcing processes. Another solution includes personalization provision and topic/domain labeling tailored according to the users' information needs. The authors in [125] proposed several solutions which when combined with classic artificial intelligence and real-time data mining methods lead to a new form of social networking, which can serve as a qualitative and credible medium of information exchange.

## 7. Conclusions

In this paper, we have reviewed two major aspects of OSNs, namely the online social influence and the role of semantics in OSNs (Section 3 and 4 respectively), while in Section 5 we discussed how we may combine both aspects towards the qualitative assessment and modeling of user-generated content. To perform a more detailed analysis and to adequately cover all perspectives of the aforementioned aspects, we analyzed the reviewed works according to a proposed hierarchical classification scheme (Figure 1).

All of the related studies regarding influence measurements in OSNs conclude that the number of followers/friends a user has, does not necessarily guarantee high influence, despite affecting it to a certain extent. The most important factors that affect users' influence can be categorized as:

- **User-oriented**: interaction with other users and similar activities (e.g. creating new messages), relationship details (number of followers, following users, friends, and so on), as well as structural network characteristics and attributes (e.g. position, shortest paths, closeness, eccentricity, centrality, and degree).
- **Content-oriented**: viral content (e.g. hashtags, mentions), identified users' interests.
- **Quality-oriented**: where quality is measured by the user's social acknowledge and the degree of engagement with other users (e.g the number of retweets/shares, favorites/likes, replies and so on).

Usually, the users are highly influential mainly on specific topics and less on others. Therefore, we found that two types of influence exist; a topic-specific one, and a global one spreading through the whole network. Several recent studies propose that social influence should be incorporated into recommendation systems to leverage past behavior and latent relationships among users, as well as to improve their performance. In parallel, social semantics are exploited in the analysis of users' behavior, interests and preferences, so as to help recommenders to suggest informative content, similar users, and other personalized information and others.

The literature that we have reviewed in this work has confirmed that influence and information flow are two interdependent concepts of OSNs, since they affect one another positively or negatively. Studies on dissemination of information have shown that the largest cascades tend to be generated by influential users who have many followers. Usually a large number of those -not so highly influential- followers initiate short diffusion chains which quickly merge into a large single structure. The dynamics of that information flow can be quantified by considering the following social diversities:

- User activity or passivity.
- User influence and susceptibility.
- User relationships in terms of interaction (e.g. mentions, replies) and friendship features.
- Reposting characteristics (e.g. volume, speed, time interval, number of hops).
- Homophily and entity similarity.
- Network attributes, structure and user topology.
- Content and structure of messages (e.g. topics, presence of URLs or hashtags, formality of language).

In addition, the study of social information spread (diffusion and propagation) is intrinsically connected to the problem of analyzing the modular structure of networks, known as community detection. Communities in OSNs promote certain topics and can be treated as the logical grouping of social actors that share common interests, ideas, or beliefs. There are two possible sources of information which can be used towards their detection: the network structure and the features and attributes of the nodes-users.

Finally, OSNs users often create messages that are characterized by the highly unstructured and informal language with many typographical errors, lack of structure, limited length, and high contextualization. Consequently, microblogging retrieval systems suffer from the problems of data sparseness and the semantic gap. To overcome those limitations and to contextualize the semantic meaning of microblog content, recent studies ([62], [71], and [117]) focus on exploiting the use of social semantics and user-generated content by identifying entities in them. These entities are used as keywords to indicate the topics of the messages, description of real-life events, as well as to reveal behavioral patterns and building interest profiles, thus enabling the interrelation of semantically related terms and the social proximity or similarity between profiles and interests. Often, those entities are linked to knowledge bases (e.g. Google Knowledge Graph, DBpedia) or are represented as concepts extracted from ontologies using Semantic Web vocabularies in order to transform unstructured data into Linked Data.